\documentclass[sigconf]{acmart}

\AtBeginDocument{%
  \providecommand\BibTeX{{%
    \normalfont B\kern-0.5em{\scshape i\kern-0.25em b}\kern-0.8em\TeX}}}

\setcopyright{acmlicensed}
\copyrightyear{2026}
\acmYear{2026}
\setcopyright{cc}
\setcctype{by}
\acmConference[CHI '26]{Proceedings of the 2026 CHI Conference on Human Factors in Computing Systems}{April 13--17, 2026}{Barcelona, Spain}
\acmBooktitle{Proceedings of the 2026 CHI Conference on Human Factors in Computing Systems (CHI '26), April 13--17, 2026, Barcelona, Spain}
\acmPrice{}
\acmDOI{10.1145/3772318.3790361}
\acmISBN{979-8-4007-2278-3/2026/04}
\usepackage[raggedrightboxes]{ragged2e}
\usepackage{tabularx}
\usepackage{enumitem}
\usepackage{booktabs}
\usepackage{multirow}
\usepackage{pifont}
\usepackage{listings}

\usepackage{lipsum}
\usepackage{ifthen}

\newenvironment{addmargin}[2][\empty]{%
  \par
  \setlength{\parindent}{0pt}%
  \rightskip=#2\relax
  \ifx\empty#1\relax
    \leftskip=\rightskip
  \else
    \leftskip=#1\relax
  \fi
}{\par}
\usepackage{xcolor}
\usepackage[normalem]{ulem} 
\lstdefinestyle{gptprompt}{
    backgroundcolor=\color{lightgray!20}, 
    basicstyle=\ttfamily\footnotesize,
    frame=single,                        
    breaklines=true,                     
    columns=fullflexible                 
}

\newcommand{\new}[1]{\textcolor{black}{#1}}
\newcommand{\IncongruencyAlt}{\mathit{Incongruency}_{\mathrm{Alt}}}
\newcommand{\IncongruencyAltNorm}{\text{Incongruency}_{\mathrm{Alt}}}

\begin{document}

\title{Mapping the Spiral of Silence: Surveying Unspoken Opinions in Online Communities}
\author{Dora Zhao}
\email{dorothyz@stanford.edu}
\affiliation{%
  \institution{Stanford University}
  \city{Stanford}
  \state{California}
  \country{USA}
}

\author{Diyi Yang}
\email{diyiy@cs.stanford.edu}
\authornote{Both authors co-advised.}
\affiliation{%
  \institution{Stanford University}
  \city{Stanford}
  \state{California}
  \country{USA}
}

\author{Michael S. Bernstein}
\email{msb@cs.stanford.edu}
\authornotemark[1]
\affiliation{%
  \institution{Stanford University}
  \city{Stanford}
  \state{California}
  \country{USA}
}

\renewcommand{\shortauthors}{Zhao et al.}

\begin{abstract}
We often treat social media as a lens onto society. How might that lens distort the popularity of political and social viewpoints? We examine discrepancies between publicly posted and privately surveyed opinions within communities, contributing a measurement of the ``spiral of silence'' theory; the theory posits people are less likely to voice opinions when they believe they hold minority views, creating a reinforcing cycle where these opinions are expressed less. We surveyed members of politically-oriented Reddit communities about their willingness to post on contentious topics, yielding 439 responses across twelve subreddits. 72.1\% of participants who perceive themselves in the minority remain silent and are half as likely to post compared to those who believe their opinion is in the majority. Community design factors, such as perceived diversity, are associated with less self-silencing. We provide recommendations for counteracting self-silencing at the community level (e.g., positive reinforcement, more transparent moderation). Overall, these results reveal gaps between online discourse and broader public opinion.
\end{abstract}

\begin{teaserfigure}
    \centering
    \includegraphics[width=\textwidth]{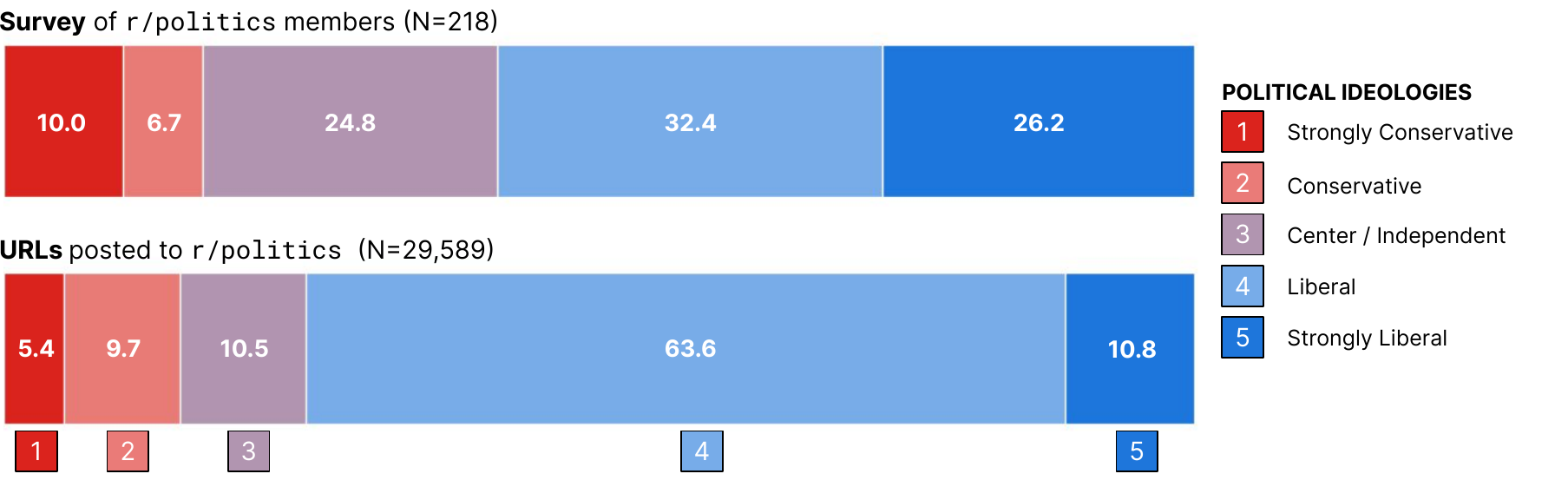}
    \caption{The political orientation of content posted on \texttt{r/politics} is not representative of its members' ideologies. Since submissions on \texttt{r/politics} must contain URLs to articles from approved news and media organizations~\cite{politicsRules}, we analyze the political leanings of the linked-to news sources. Compared to the surveyed political orientations of \texttt{r/politics} members with the ideologies expressed in the links shared to the subreddit, there is $1.3$ times more liberal-leaning links than active liberal members on \texttt{r/politics} (see Appendix~\ref{sec:teaser} for methodological details).}
    \label{fig:teaser}
    \Description[Comparison between the political orientation of content posted on r/politics and the surveyed political ideologies of its active members from Prolific.]{Comparison between the political orientation of content posted on r/politics and the surveyed political ideologies of its active members from Prolific. The figure shows that 74.4\% of shared links come from liberal-leaning sources, while only 58.6\% of active members identify as liberal, revealing a distortion in representation.}
\end{teaserfigure}



\maketitle
\section{Introduction}
Content shared on social media is often treated as a reflection of broader public opinion~\cite{anstead2015social,mcgregor2020taking,asur2010predicting}. But is this view distorted? For example, in 2019, a racist photo of the governor of Virginia surfaced online. As reported in \textit{The~New~York~Times}, this event led to universal condemnation of the governor on social media, leading major politicians to demand that he resign~\cite{cohn2019dem}. However, a poll of Virginians showed that most people, and especially Black constituents, were more in favor of the governor remaining in office~\cite{jamison2012wapo}. Such voices were less vocal and less visible on social media, despite their prevalence.

Divergence between what we see on social media and what the public believes is not uncommon. For example, comparing the content submitted to the subreddit \texttt{r/politics} --- a massive online community for sharing and discussing US political news ---  to the surveyed ideologies of active members, there is a large distortion (details in Appendix~\ref{sec:teaser}): $74.4\%$ of the content shared on the subreddit has a liberal view, compared to only $58.6\%$ of what active members self-report (Fig.~\ref{fig:teaser}). This distortion can arise in liberal, conservative, and mixed spaces. What causes this distortion? While this could be an artifact of users' preferences for posting online on certain topics, or perhaps a product of social loafing~\cite{karau1993social}, we explore an alternative: users may refrain from sharing their opinion when they perceive themselves as being in the minority. The spiral of silence theory~\cite{noelle1974spiral} explains this behavior as a response to the fear of social isolation, where individuals self-silence --- or decide not to share their opinion --- rather than risk social disapproval. Self-silencing then leads to less of the opinion being expressed in the group, making the viewpoint appear even further minoritized. This reaction produces a downward spiral as people become more likely to self-silence. Since being proposed, the spiral of silence has been empirically studied in face-to-face communication across multiple issue types and participant populations~\cite{scheufle2000twenty,matthes2018spiral}. As online communication forms like social media proliferate, it raises the question of how the spiral of silence might affect user behavior in this setting.

Measuring the spiral of silence on social media, especially across diverse online communities, is challenging. Not all topics are likely to trigger the spiral of silence~\cite{lee2004cross}, and identifying such topics may require insider knowledge of community norms and practices~\cite{chen2013and,hessel2019something}. Furthermore, even with a set of topics in hand, the process of measuring rates of self-silencing requires capturing user attitudes, beliefs, and behaviors that cannot be observed from online content itself. This traditional approach of analyzing existing posts online is self-defeating, since it excludes viewpoints that users feel uncomfortable sharing online.

In our work, we develop a human-plus-algorithm pipeline to generate potential topics that community members feel are appropriate and in-bounds for a community to discuss, but which also likely to spark internal disagreement. Then, drawing on survey methodologies common in empirical studies of the spiral of silence~\cite{matthes2018spiral}, we measure whether the spiral of silence occurs on those topics in those communities. Further, we evaluate whether online community factors such as moderation and diversity mitigate or amplify the spiral. We apply this method to study the spiral of silence on Reddit --- a popular social media platform in the United States --- across a number of subcommunities (``subreddits'') focused on political and social issues but with varying norms and values. Importantly, our analyses focus on topics that community members agree are acceptable to discuss within the subreddit but are still hesitant to engage with, and exclude topics considered inappropriate or in violation of community norms.

We find that participants who believe their opinion is in the minority remain silent $72.1\%$ of the time, and these opinions are only half as likely to be posted compared to those in the majority. Participants are also less likely to upvote posts sharing minoritized opinions. However, we identify two community design factors that are linked to variations in the rate of self-silencing. We test these results through a series of three mixed-effects models, finding that participants report a higher likelihood of sharing minoritized opinions in subreddits they perceive as more diverse, but a lower likelihood of sharing in subreddits with more stringent content moderation.\footnote{Code available at \url{https://github.com/StanfordHCI/spiral-of-silence}.}

Our findings illustrate how, within the politically-oriented subreddits we study, the distribution of viewpoints being shared online can misrepresent community members' actual opinions, systematically marginalizing minority perspectives. There are certainly some topics where minority opinions are normatively undesirable to represent, even if a portion of the population agrees with them, such as, hate speech or misinformation. However, for many other topics where the minority voice instead reflects under-represented groups or moderate views from people who believe themselves to be in the minority, communities may want to mitigate the spiral and hear from a more representative set of voices~\cite{farjam2024social,bowen2003spirals}. Based on our findings, we surface tangible actions that online moderators and platform designers can take, such as providing positive feedback on posts expressing diverse viewpoints and increasing moderation transparency, to help mitigate the spiral of silence.

\section{How the Spiral of Silence Shapes Social Media}
Social media is often viewed as a mirror of society, or at the very least, one of our best available instruments for understanding public opinion. Social media data has been used for a range of tasks from forecasting box-office performances~\cite{asur2010predicting} to understanding political support for candidates~\cite{mcgregor2020taking}. Furthermore, social media content can actively shape people's opinions and influence their actions~\cite{laranjo2015influence}.

However, what we see online may not be a complete or accurate picture of public opinion. This is partly because online content originates from a small fraction of platform users~\cite{ruths2014social}. The majority of online communities consist of ``lurkers'' or individuals who visit the community but choose to remain silent~\cite{sun2014understanding}. There are many reasons why people may choose to stay as lurkers. In some cases, users may just want to learn about a topic or are not motivated to post if they see their opinion reflected in existing content~\cite{nonnecke2001lurkers}. Other times, however, users may want to share their opinions but are uncomfortable doing so. The disproportionate silencing of viewpoints may make our online environments appear more polarized than they actually are.

One explanation for this phenomenon is \citet{noelle1974spiral}'s spiral of silence theory. The theory posits that, when people perceive their viewpoints as being \textit{incongruent}, or different from the majority of the group, they are less likely to engage in \textit{opinion expression}, or share their viewpoint. Since fewer people express the opinion, the viewpoint appears to be even further in the minority, triggering a spiral effect as the growing silence of incongruent opinions compounds upon itself.

The spiral of silence is an empirical effect, not an unavoidable constant applying to all people: there are ``avant-garde'' and ``hard core'' individuals who stand up for their viewpoints even when they are in the minority~\cite{noelle1977turbulences}. Through empirical studies,  work has proposed factors that modulate the spiral of silence. Individual-level antecedents, including personality traits~\cite{hayes2005willingness,neuwirth2007spiral}, opinion certainty~\cite{matthes2018spiral}, and issue type~\cite{gearhart2018same} can influence opinion expression. In social media settings, prior work has focused on identifying platform affordances that influence self-silencing behavior~\cite{wu2018comment,pang2016can}. For example, greater perceived anonymity can dampen the effects as the potential social consequences of speaking out are diluted~\cite{wu2018comment}. Even the means by which opinions are shared, such as commenting versus liking content, can impact the degree of self-silencing~\cite{pang2016can}.

While we have insight into how platform-level design decisions influence self-silencing, platforms can house many online communities each with their own rules, norms, and values. Given the unique configuration of each online community, it is not surprising that users' behaviors, such as conversation patterns~\cite{choi2015characterizing} or self-disclosure~\cite{yang2017self}, will differ across communities. At the moment, the impact of these community-level design decisions has not been factored into our understanding of how the spiral of silence manifests. To better understand how these design decisions are linked to self-silencing, this work focuses on Reddit, which is comprised of many distinct subreddits --- each of which has its own governance structures, moderation practices, values, topical interests, and norms~\cite{weld2021making,fiesler2018reddit,leibmann2025reddit} --- providing an ideal context for studying community-level design variation. This leads us to our research questions:

\begin{addmargin}[3em]{0em}
\textbf{RQ1.} How does users' opinion expression on political subreddits differ when they perceive their viewpoint as being in the majority versus the minority?\\
\textbf{RQ2.} How do online community design factors on political subreddits influence opinion expression of incongruent viewpoints?
\end{addmargin}

\section{Factors Influencing Opinion Expression}
\label{sec:factors}
In this section, we explore how the spiral of silence and community design decisions can shape opinion expression on social media and introduce our hypotheses.

\subsection{Opinion Incongruency}
The spiral of silence theory provides an explanation for viewpoint sharing patterns on social media platforms. Previous studies~\cite{pewresearchcenter_2014_sos,kushin2019societal,pang2016can,hoffmann2017spiral,gearhart2015something,liao2016snowden} on social media platforms, such as Facebook and Twitter, found evidence that users are less likely to express incongruent viewpoints, or opinions that differ from what they perceive the majority of other members believe. We expect a similar phenomenon on Reddit. Since subreddits also offer a sense of community and socializing opportunities~\cite{moore2017redditors}, community members may still experience a fear of social isolation. This fear can be triggered on Reddit by potential social repercussions, such as being downvoted or facing personal attacks from others in the community~\cite{neubaum2018we,cheng2017anyone}. Especially when discussing stigmatized topics, users also have expressed concerns about exposure, including having sensitive content linked to their public profile or facing offline repercussions from people who know about their online account~\cite{leavitt2015throwaway,reagle2023even}. These consequences encourage users to self-silence when they have an opinion that they think differs from the majority. Given this evidence, we expect our study will yield results in accordance with the spiral of silence: 
\begin{addmargin}[3em]{0em}
\textit{Hypothesis 1: The likelihood of opinion expression is negatively associated with viewpoint incongruency (i.e., when users believe the majority of community members disagree with them).}
\end{addmargin}

\subsection{Community Inclusion and Diversity}
How online communities are designed will influence how users behave, especially regarding sharing their opinions publicly. A greater sense of belonging can lead users to be more active as they feel more included and trusting of others in the group. Similarly, feeling psychologically safe can foster confidence and empower users to voice their opinions~\cite{edmondson2014psychological}.

Feeling a sense of belonging is an essential human need that applies not only to offline but also online settings~\cite{maslow1958dynamic,lambert2013belong}. \citeauthor{hagerty1992sense} proposed two key dimensions that define a sense of belonging: (1)~having involvement in a group be valued, and (2)~feeling a sense of fit or congruency with other group members. When these criteria are met, people are more likely to be involved in a group and find this involvement to be more meaningful~\cite{hagerty1996sense}. The same principles apply to belonging in online communities~\cite{lampe2010motivations}. Users with strong shared identities or bonds are more likely to be active in online groups~\cite{ren2007applying}. When people feel included, they are more likely to have a sense of familiarity and trust other community members~\cite{zhao2012cultivating}. This encourages users to share more freely, as they feel less at risk of being judged when sharing their viewpoints. Thus, we expect that:

\vspace{0.25em}
\begin{addmargin}[3em]{0em}
\textit{Hypothesis 2a: The perceived inclusivity of a community is more positively associated with the likelihood of opinion expression when users hold incongruent viewpoints.}
\end{addmargin}
\vspace{0.25em}

Who is part of the community also matters. Users frequently mention the fear of being personally attacked as a primary reason for not posting minority opinions~\cite{neubaum2018we}. In both online and offline settings, existing research has found that people are more secure when sharing ideas in psychologically safe communities~\cite{newman2017psychological}. That is, when people foresee fewer negative interpersonal consequences, they are more willing to take potentially controversial actions. Community diversity is an important antecedent for psychological safety~\cite{newman2017psychological}. Emphasizing diversity can foster a positive climate in which community members, regardless of their background, more strongly identify as members of the group and feel psychologically safe~\cite{mckay2007racial}. In this setting, group members are more willing to engage in potentially risky behaviors, such as sharing opinions that may deviate from what is popular. In a similar vein, prior work on the spiral of silence has suggested that those with more bridging social capital that connects them to heterogeneous networks on social media platforms, are more willing to speak up online~\cite{sheehan2015change}. From this literature, we hypothesize the following:

\vspace{0.25em}
\begin{addmargin}[3em]{0em}
\textit{Hypothesis 2b: The diversity of community members is more positively associated with the likelihood of opinion expression when users hold incongruent viewpoints.}
\end{addmargin}
\vspace{0.25em}

To conceptualize inclusion and diversity, we draw on work that has sought to quantify the values of online communities~\cite{weld2022makes}. We utilize the instrument proposed by~\citet{weld2021making} to capture users' perceived sense of inclusion and diversity across different subreddits. Perceived inclusion is defined as how ``included and able to contribute new and existing member feel'' while perceived diversity refers to how diverse people in the community are thought to be.

\subsection{Community Moderation}
Moderation practices frequently shape the dynamics of online communities. Moderators can set the tone of the online community by encouraging other users to imitate their actions and by curbing instances of behavior that they deem to fall outside community norms~\cite{seering2017shaping}. Removing or prescreening content that does not adhere to the community's norms is an effective way of reducing ``bad behavior'' and can positively impact community health~\cite{kraut2012building,seering2017shaping}. By filtering out potentially harassing or offensive submissions, moderators can assuage members' fear of being personally attacked and facilitate more open opinion-sharing~\cite{gibson2019free}. For example,~\citet{wadden2021effect} studied mental health discussions in moderated and unmoderated spaces. They found that users were not only more likely to post in moderated spaces, but also that they were more willing to share vulnerable messages in these discussions. In a setting more similar to opinion-sharing on Reddit, \citet{wise2006moderation} found that users were more willing to participate in a moderated online political community compared to an unmoderated space. From these results, we expect that users feel more comfortable sharing minoritized opinions in communities with more active moderation. Thus, we hypothesize:

\vspace{0.25em}
\begin{addmargin}[3em]{0em}
\textit{Hypothesis 3: Moderation activity is more positively associated with the likelihood of opinion expression when users hold incongruent viewpoints.}
\end{addmargin}
\vspace{0.25em}

In this work, we focus on moderation through content removal. We acknowledge that content removal represents only a subset of moderation actions available on Reddit, but choose to focus on content removal because it is one of the most common moderation interventions and has been shown to influence community members' participation~\cite{li2022all}.

\section{Method}
\label{sec:methods}
Our goal is to measure the spiral of silence across topics and communities. We first need a set of controversial topics that may trigger the spiral of silence. This analysis requires us to identify topics that community members agree are permissible to discuss in their community in principle, but may not feel comfortable sharing in practice. Crawling existing posts from these communities, unfortunately, is insufficient: these posts will not reflect the opinions that users may be hesitant to share in the first place. In other words, starting with existing topics runs the risk of only identifying topics that are so safe as to appear commonly in the community. Alternatively, while asking participants to list instances when they have self-silenced is possible, this elicitation process may be prone to recall or desirability bias and yield idiosyncratic or off-scope topics falling outside the theory’s bounds.

Thus, we use a hybrid human-AI method that draws on large language models (LLMs) to nominate potential controversial topics that are appropriate to each community, even if those topics are not posted in the community, and validate those proposed topics with active subreddit members. We then survey community members to capture the rate of self-silencing on those topics. In this section, we start by discussing the scope of our study and our definition of self-silencing. Then, we provide an overview of our method, describing the topic generation process and survey instrument. This study was approved by Stanford University's Institutional Review Board.

\paragraph{Scope of the Study} One important distinction we make in this work is between when community members self-silence because they are uncomfortable sharing an opinion versus when an opinion is not considered acceptable to be shared within a community. Our study is constrained only to the former. In other words, we focus only on perspectives that lead to disagreement while still falling within the bounds of the community norms. We do not seek to make any normative claims about what should or should not be shared to a subreddit. Instead, we are interested in what is not being shared, even when most community members agree the opinion \emph{should} be allowed to be discussed.

\subsection{Topic Generation}
\label{sub:topic-generation}
\begin{figure*}[t]
    \centering
    \includegraphics[width=\textwidth]{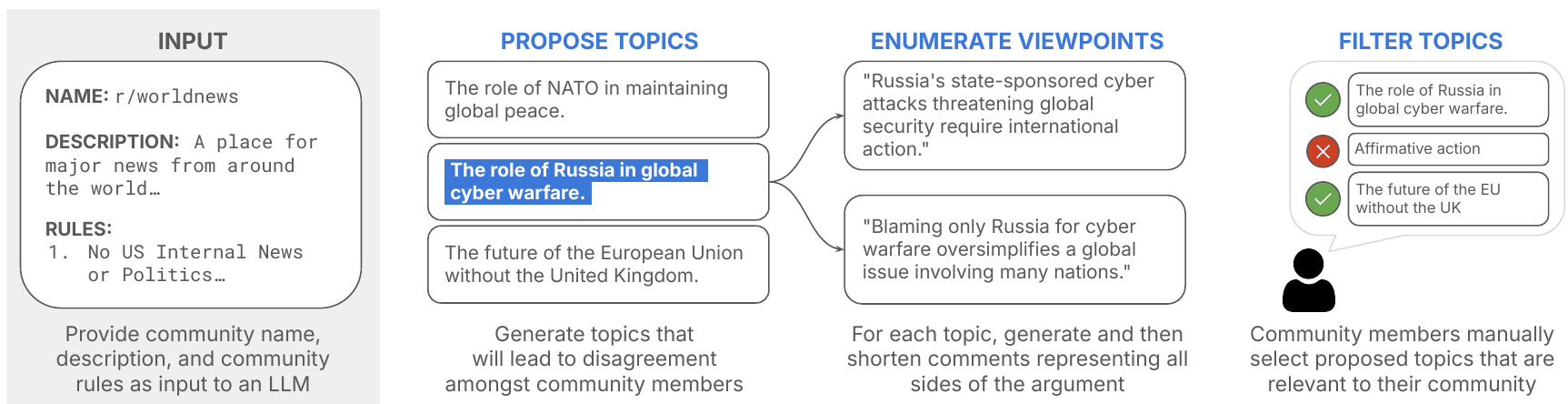}
    \caption{We generate controversial viewpoints specific to a provided community. Using the community name, description, and rules, we prompt an LLM to propose controversial topics. For each topic, we enumerate statements representing different stances on the topic. Finally, the generated topics and viewpoints are validated by community members to ensure they are relevant and fall within the norms of what is accepted in the community. }
    \label{fig:generation_method}
    \Description[Overview of the topic generation pipeline.]{Overview of the topic generation pipeline. The diagram illustrates how subreddit information (name, description, and rules) is provided to an LLM, which generates potentially controversial topics and opposing viewpoints. In the final step, community members then validate these topics to ensure relevance and appropriateness.}
\end{figure*}

Our methodological approach involves using an LLM to propose controversial topics and opposing viewpoints for each topic, then surveying actual community members to select viewpoints for our study, which provides several advantages for studying the spiral of silence in our setting. First, while manually selecting topics is the de facto approach when studying the spiral of silence across more than one issue type~\cite{gearhart2018same,lee2004cross}, manual selection may introduce experimenter bias and may be impractical across a large number of diverse yet specific communities on Reddit, each with its own rules and norms~\cite{fiesler2018reddit}. Deciding what topics are controversial will be imprecise for those without specific knowledge of the community~\cite{hessel2019something,chen2013and}. Second, directly looking at Reddit data is not adequate in this scenario either, as we are interested in the opinions that may be less likely to be shared to the community, rather than those that are already present. Thus, using mined data from online only gives us a window into the viewpoints that community members feel comfortable sharing. By contrast, using an LLM allows us to nominate potential topics that are relevant to the community and might result in self-silencing, including perspectives that may be underrepresented or missing in the current discourse online. Prior work has also shown that LLM-based topic generation produces more coherent and interpretable topics than traditional approaches such as LDA~\cite{pham2024topicgpt,kaur2024moving}. We then ask active community members to validate the relevancy and appropriateness of these proposals, selecting topics from this filtered set to include in our study.

\subsubsection{Proposing Controversial Topics and Viewpoints} Topic generation is a three-step process (see Fig.~\ref{fig:generation_method}). We provide a community name, description, and rules as input to an LLM (\texttt{GPT-4}) to propose 20 potentially controversial topics for the provided community. We provide in our prompt that we define ``controversial'' as being likely to cause disagreement between community members. Then, for each proposed topic, the LLM generates two viewpoints representing different sides of the issue using few-shot prompting techniques (see Appendix~\ref{sec:hai_design} for prompts). The viewpoints must follow the rules of the subreddit, which are again provided to the model. Finally, as we discuss in detail in Sec.~\ref{sec:valid_generate}, viewpoints are manually validated to ensure that they are comprehensible and relevant to the respective subreddit. 

\subsubsection{Selecting Political Subreddits}
\label{subsec:sublist}
To run our topic generation pipeline, we must select which subreddits to provide as input to the LLM. \new{For this work,} we focus on subreddits that focus on political and social issues, since \citet{noelle1974spiral} posits that morally laden subjects typically trigger the spiral of silence. To identify these communities, we start with the 2,040 subreddits on \texttt{r/ListOfSubreddits} and sample the 50 hottest submissions in each as of January 2024 accessed via the Reddit API. We then filter for the following criteria: (1)~the majority of submissions are in English as classified by \texttt{langdetect}~\cite{nakatani2010langdetect}; (2)~$\geq80\%$ of submissions are classified to contain political content using an LLM classifier on the post title following an approach from prior work~\cite{piccardi2024social} (see Appendix~\ref{sec:app_subreddit_selection} for prompt); and (3)~the subreddit has over 100K members. Filtering leaves us with 23 political subreddits.

Since we seek to compare variation in opinion expression on topics based on community-level differences, we want to identify a single, static set of topics that are likely to be relevant across all 23 subreddits. To ensure broad applicability of the topics, we selected \texttt{r/politics} and \texttt{r/worldnews} as input to the LLM for topic generation based on their size and generality, rather than using more specialized subreddits that might yield niche topics with limited relevance across communities. Using our pipeline, we generate 40 topics (i.e., 20 per subreddit) with opposing viewpoints for each topic. As shown in Table~\ref{tab:shortened_list}, topics covered a range of political and social issues, including abortion rights, military spending, and affirmative action. Each identified \textit{topic} contains two \textit{viewpoints} that articulate two main clusters of opinion on the topic: for example, with affirmative action, one viewpoint states, ``\emph{I believe affirmative action counters systemic biases and fosters a diverse, inclusive society},'' and the other viewpoint states, ``\emph{I believe affirmative action could unintentionally
cause reverse discrimination and undermine merit,
potentially increasing societal division}.'' For the full list of topics and viewpoints, refer to Appendix Table~\ref{tab:viewpoint_list}.

\subsection{Validating Generated Topics and Viewpoints}
\label{sec:valid_generate}
Once we have our generated topics, how do we ensure that they fall within the bounds of what is acceptable for a given subreddit? We conduct three checks to validate and filter our generations. First, we manually inspect each of the generations for coherence. Then, we survey active community members to certify that the generated topics are relevant and appropriate to the subreddit. Finally, we conduct an ``intrusion test'' also with active community members to verify that our generated topics could plausibly appear in posts on the subreddit. We apply our checks in a sequential order, meaning that only the topics that pass the coherence check are included in the community-specific relevance check and so on.

\subsubsection{Coherence Check} \new{We start by
manually filtering the generations for comprehension.} Using the 40 topics from the prior step, we check \new{that the generated outputs contained opposing viewpoints, covered controversial topics, and were worded in a manner that participants would understand. From this step, we filtered out seven topics:} four had incomplete viewpoints (i.e., the model only provided viewpoints representing one side) and three were either worded confusingly or presented viewpoints that did not present opposite stances, leaving us with 33 topics. 

\subsubsection{Community-Specific Relevance Check} 
\label{sec:community-check}
Next, after ensuring that generations are coherent, we need to validate that our generated topics and viewpoints are not only relevant, but also follow the community norms for the respective subreddit. The first way we do this is by having active members of the subreddits (\texttt{r/politics} and \texttt{r/worldnews}) manually review the topics. We recruited Prolific participants who self-reported as active members. All participants first completed a screening quiz to confirm their knowledge about the subreddits in which they were presented a set of three post titles---two from their selected subreddit and one from another political subreddit (which we manually verify is irrelevant to the selected subreddit)---and asked to select which titles were likely to appear in their selected subreddit (see Appendix~\ref{sec:app_kq} for more details and examples). If they identified all the correct post titles, participants qualified for the survey. In total, 46 participants passed our screening quiz and completed our community-specific topic relevance check. They were then shown a list of generated topics and viewpoints and asked whether they were relevant to the subreddit. Participants also had the option to response ``I don't know'' for topics. We removed two topics that many respondents reported not having knowledge on and one topic that was commonly marked as not relevant to the community. Ultimately, from this filtered set of 30 topics, we randomly selected 11 for our final study --- five generated with \texttt{r/politics} as the input and six with \texttt{r/worldnews} --- to ensure there are an adequate number of responses per topic for our analysis.

\subsubsection{Intrusion Testing}
\label{sec:intrusion}
While our coherence and community-relevance checks ensure generated topics are understandable and broadly aligned with subreddit norms, they do not tell us whether these topics \emph{feel} authentic to active community members. A topic may be technically relevant but still strike community members as out of place. To examine whether our topics meet this bar, we conduct an intrusion test, a method of outlier detection widely used in natural language processing as a quantitative evaluation for topic modeling~\cite{chang2009reading,bhatia2018topic,churchill2022evolution}.

\paragraph{Task Setup} We recruited a new set of active community members (i.e., Prolific participants who passed our screening quiz for the respective subreddit) to participate in our intrusion testing task. These participants were shown the generated topic and four controversial topics mined from the subreddit that are semantically similar to the generated topic. To obtain the controversial topics mined from the subreddit, we start with Pushshift dumps of all posts made in \texttt{r/politics} and \texttt{r/worldnews} between January 2020 and December 2023~\cite{pushshift2023dumps}. Then, following the pipeline from \citet{hessel2019something}, we first discard posts that have less than 30 comments and sort posts by their upvote ratio. After removing posts with a negative score, we randomly select 500 posts in the bottom quartile of scores, which contain topics that subreddit members want to engage with (as indicated by the number of comments under the post) but lead to high amounts of disagreement (as indicated by the close number of upvotes and downvotes.). Finally, we extract common controversial topics using TopicGPT~\cite{pham2024topicgpt}. For each generated topic, we select the four most semantically similar mined topics as determined by the cosine similarity between sentence embeddings~\cite{reimers2019sentence}. For example, 
for the generated topic of ``abortion rights'', the mined topics were ``impact of Roe v. Wade's overturning'', ``Amy Coney Barett's confirmation'', ``freedom of protest rights'', and ``government regulation of personal freedoms and public health.'' Participants then identified the ``intruder'' topic, by selecting which of the five topics is least likely to appear in the subreddit. If our generated topic is inappropriate to a subreddit, then it should be consistently selected as the intruder. 

\paragraph{Results} In total, 63 participants completed our task -- 32 from \texttt{r/politics} and 31 from \texttt{r/worldnews}. Across the eleven topics, our generated topic was selected as the intruder only 18.8\% of the time. A one-proportion $z$-test comparing this intrusion detection rate against random guessing (20.0\% selection rate) was not statistically significant ($z = -0.41$, $p = 0.68$), indicating that community members were no better than chance at identifying our generated topic. This result suggests that the generated topics are likely to appear in the subreddit according to community members, and for nine of the eleven topics, they are rated as more likely to appear in the subreddit than topics from mined conversation data.

\begin{table*}[]
    \centering
    \small
    \caption{Examples of topics, and their respective viewpoints, generated for \texttt{r/politics} and \texttt{r/worldnews}.}
    \begin{tabular}{
    >{\raggedright\arraybackslash}p{2.5cm}
    >{\raggedright\arraybackslash}p{5.2cm}
    >{\raggedright\arraybackslash}p{5.2cm}
    }
    \toprule
    \textbf{Topic} &  \textbf{Viewpoint 1} &  \textbf{Viewpoint 2}\\
    \midrule
    Universal Healthcare & I believe in Universal Healthcare because everyone deserves access to good health, funded by the government. & I believe market competition and individual insurance plans are superior to Universal Healthcare.\\
    \addlinespace
    Abortion Rights  & As a pro-choice supporter, I believe women's bodily autonomy and reproductive choices are crucial for gender equality. & As a pro-life advocate, I believe every life from conception deserves legal protection.\\ 
   \addlinespace
    Drone warfare & 
    Drone warfare is a necessary evil for global security due to its precision, efficiency, and safety for soldiers.
    & 
    Drone warfare inevitably causes collateral damage, violates human rights, and induces terror.\\
    \bottomrule
    \end{tabular}
    \label{tab:shortened_list}
    \Description[Examples of topics and opposing viewpoints generated for r/politics and r/worldnews using the topic-generation pipeline.]{
    Examples of topics and opposing viewpoints generated for r/politics and r/worldnews using the topic-generation pipeline. For each topic, the table presents two representative statements capturing distinct perspectives. The example topics include universal healthcare, abortion rights, and drone warfare.
    }
\end{table*}

\subsection{Opinion Expression Survey}
\begin{figure*}[h!]
    \centering
    \includegraphics[width=\textwidth]{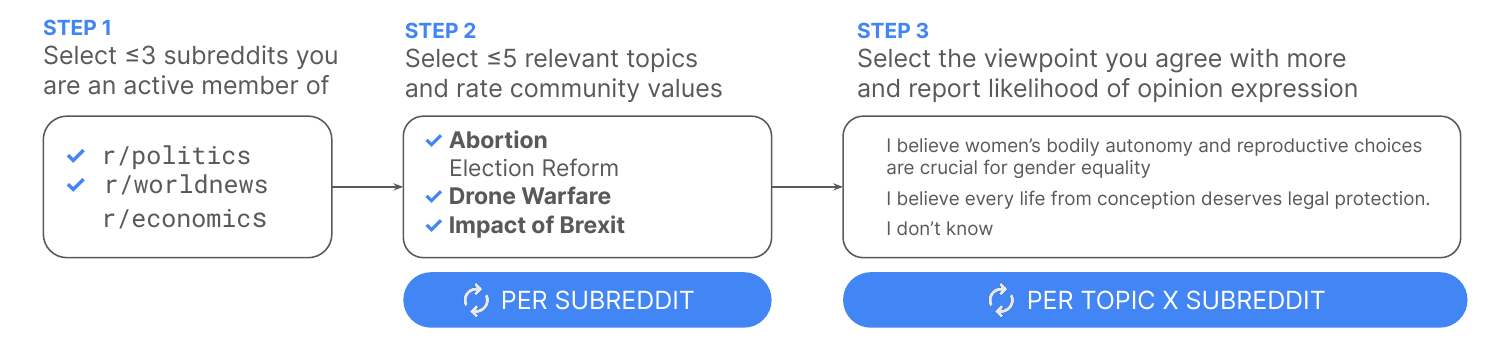}
    \caption{We measure self-silencing behavior across topics and subreddits using a survey. First, participants select up to three subreddits of which they are an active member. For every selected subreddit, the participant answers questions about the community's values and chooses up to five topics relevant to the community. For each topic, we ask the participants which viewpoint they agree with more. Finally, for the viewpoint that they agree with, participants answers how much they agree with the viewpoint, how much they believe others in the subreddit agree with the viewpoint, and their likelihood of sharing the viewpoint.}
    \label{fig:survey-flow}
    \Description[Overview of the opinion expression survey.]{Overview of the opinion expression survey. The diagram illustrates the steps in our survey. Users first select subreddits that they are an active member of. They then select topics that are relevant to that subreddit. For each topic, they select which viewpoint they agreed with.}
\end{figure*}
Next, we use the validated viewpoints from Sec.~\ref{sec:valid_generate} to gauge the rate of self-silencing among active community members.\footnote{Our study is pre-registered on OSF at this \color{ACMDarkBlue}\href{https://osf.io/gze65/?view_only=03f45025105a42b1b50cd6936ece06d0}{link}.} Since these self-silenced viewpoints will not be posted on Reddit, we directly survey community members to measure this phenomenon. The goal of our survey is to identify how willing an individual is to share their opinion on a topic to a subreddit. Specifically, we are interested in the following dimensions: (1) differences in sharing behavior when individuals view themselves as being in the majority versus minority; and (2) how self-silencing differs across subreddits.

\subsubsection{Survey Instrument}
For each topic, our survey captures the likelihood that participants would share their viewpoint on each subreddit. As shown in Fig.~\ref{fig:survey-flow}, participants are first asked to select up to three subreddits of which they are an active member. For each of the subreddits they selected, participants chose up to five topics they thought were relevant to each subreddit from the total pool of eleven topics. We include this step as an additional check --- on top of the validation described in Sec.~\ref{sec:valid_generate} --- to ensure that we are surveying participants about topics they believe are appropriate and relevant to the community. For example, for \texttt{r/combatfootage}, selected topics included ``ethics of drone warfare in the Middle East,'' ``military spending,'' and ``the role of NATO in maintaining global peace'' whereas in \texttt{r/feminism}, relevant topics included ``abortion rights'' and ``universal healthcare.'' 

For each topic, participants indicated the viewpoint they agreed with. They then rated the extent to which they agreed with the viewpoint, how much they believed the majority of subreddit members agreed with the viewpoint, and whether they thought the viewpoint should be allowed to be posted on the subreddit. For all viewpoints the participant agreed with and thought should be allowed on the subreddit, we asked about their likelihood of sharing that viewpoint on the subreddit and their likelihood of upvoting that viewpoint if someone else shared it on the subreddit. \new{If the participant responded that they were unwilling to comment their opinion, we asked what alternative actions (e.g., read discussion but not post, share to a different subreddit) they would take.} The order in which the topics were presented for each subreddit was randomized.

\subsubsection{Participants} 
Survey participants were recruited via Prolific. To qualify, participants were required to live in the United States, be at least 18 years old, and identify as an active member of at least one political subreddit included in our list. This list contains \texttt{r/politics} and \texttt{r/worldnews} as well as the 21 additional political subreddits from Section~\ref{subsec:sublist}. We include these additional subreddits to obtain a larger pool of participants.

Participants selected up to three subreddits they were an active member of from the list of 23 subreddits used in the pre-screener. Participants could, if desired, provide up to one additional political subreddit not included in our list. We verify participants are active members of the subreddit using the same knowledge quiz on identifying relevant post titles from the topic generation process. A total of 489 participants completed the pre-screener; 337 self-reported as members of a listed subreddit; and 290 passed the knowledge quiz.

\new{We conduct a Monte-Carlo power analysis to determine the sample size required to detect significance using a $z$-test for the fixed effect of \emph{Incongruency} at $\alpha=0.05$ and assuming a coefficient of $\beta=-0.6$}. Based on this power analysis, we needed to collect 270 responses (participants $\times$ topics $\times$ subreddit) to achieve $90\%$ power, which is approximately 54 participants. In our final analysis, we include \textbf{439 responses} from 58 participants, covering twelve subreddits.

\subsubsection{Data Collection Process}
\begin{figure*}
    \centering
    \includegraphics[width=0.98\linewidth]{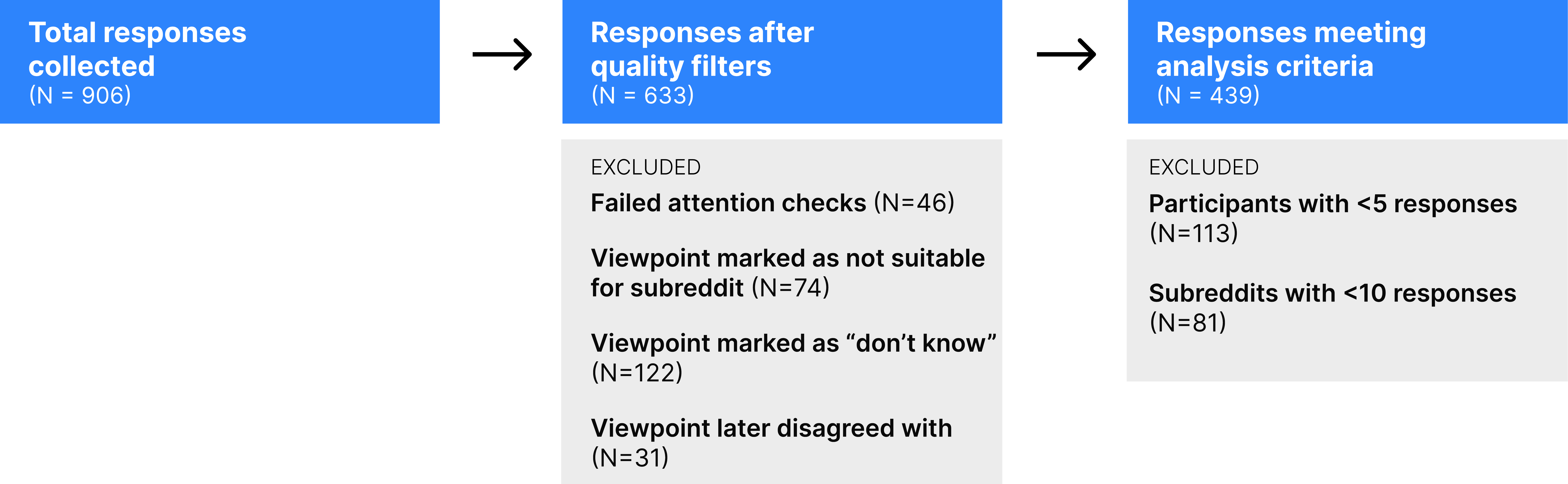}
    \caption{We filter the responses --- where an individual response denotes a participant's likelihood of speaking out on a specific topic in a subreddit ---  collected to those that pass our quality checks and meet our analysis criteria. From our initial pool of 906 responses, our filtering process yield in 439 responses that we use in our analyses.}
     \Description[Flowchart of the data filtering process for survey responses.]{Flowchart of the data filtering process for survey responses. Starting from 906 collected responses, successive filtering steps include removing failed attention checks, viewpoints marked as not suitable for a subreddit, viewpoints where participants marked "I don't know", and contradicting viewpoints (i.e., participant first agreed with a viewpoint and then said they disagreed). Then, participants with < 5 responses and subreddits with < 10 responses are filtered.}
    \label{fig:flow}
\end{figure*}

To arrive at this sample, we initially collected 906 responses from 125 participants and after filtering arrive at our final sample of 439 responses (see Fig.~\ref{fig:flow}). First, since we focus on political topics, we recruited a balanced sample of participants by political leaning. We asked for participant's political leaning in our screening quiz and stratified recruitment, resulting in 40 Democrats, 38 Republicans, 40 Independents, and 7 participants with other political ideologies. Then, for quality assurance, we removed 46 responses from 10 participants who failed our attention checks. We also removed individual responses if: (1)~the participant felt the viewpoint should \emph{not} be shared on the subreddit (N=74); (2)~the participant selected ``don't know'' when asked to pick a viewpoint for a topic (N=122); or (3)~the participant later disagreed with the viewpoint they selected as representing their stance (N=31). Finally, due to the observed sparsity in our crossed random effects structure, we restricted the dataset to subreddits with $\geq 10$ responses and participants with $\geq 5$ responses based on selections from prior work to ensure the mixed-effects models we use in our analysis have stable estimates~\cite{maas2005sufficient}. Over half of the subreddits from the initial responses had only one participant responding, which can lead to convergence issues and unreliable variance estimates in mixed-effects models~\cite{bell2008cluster}. To verify that this filtering did not introduce systematic bias, we conducted statistical tests comparing the filtered and unfiltered datasets, finding no significant differences in the distributions (see Appendix~\ref{sec:app_comparing} for full results). This process results in our final dataset of 439 responses from 58 participants, who consist of 21 Democrats, 21 Republicans, 14 Independents, and 2 with other political ideologies.

\subsection{Measures}
\label{sec:measures}
We detail the measures used in our study. Our dependent variable is the likelihood of opinion expression. We introduce four independent variables: opinion incongruence, community inclusion, community diversity, and content removal rate. Finally, we include controls for individual-level factors that can influence opinion expression.

\subsubsection{Likelihood of Opinion Expression}
Our main dependent variable is a participant's likelihood of opinion expression. In our survey, we follow prior work~\cite{zerback2017can,liu2017people,neubaum2018we} and present the participants with a hypothetical scenario. Participants are asked to imagine they see a post on their selected subreddit related to the presented controversial topic, and that no comments with the following viewpoint have been raised yet. Then, we ask their likelihood of sharing the viewpoint, \textit{Share Likelihood}, under their main account reported on a scale from 1 to 7 (i.e., ``Rate the likelihood you would share this viewpoint on the selected subreddit under your main account''). We also ask participants to provide the likelihood of upvoting the viewpoint (\textit{Upvote Likelihood}) on a seven-point scale (i.e., ``Rate the likelihood you would upvote someone else’s comment expressing this viewpoint, if the comment was present''). 

\subsubsection{Opinion Incongruency}
For a given viewpoint, \textit{Incongruency} is a binary variable indicating whether a participant believes their opinion is held by the majority~(0) or minority~(1) of other subreddit members. Following prior work~\cite{neuwirth2007spiral,chia2014authoritarian}, we operationalize incongruency as the difference between how much a participant agrees with a viewpoint and how much they believe the majority of subreddit members agree. Both agreement scores are measured on a 7-point scale. Consistent with prior work that empirically measures the spiral of silence~\cite{gearhart2018same,matthes2010spiral,neuwirth2007spiral}, we then convert this difference into a binary variable. We indicate all instances when the participant agrees with the viewpoint, and they believe the majority of subreddit members also agree as 0. Conversely, if the participant agrees with the viewpoint but believes the majority of subreddit members are neutral or disagree, incongruency is marked as 1.\footnote{As a robustness check, we report results using an alternative dichotomization for Incongruency, finding qualitatively similar results (see Appendix~\ref{sec:app_altinc}).}

\subsubsection{Community Inclusion and Diversity}
We collect self-perceived community inclusivity and diversity from participants using~\citet{weld2022makes}'s instrument for measuring community values across subreddits. For a given subreddit, participants report their perception of the current state of \textit{Inclusion} (i.e., ``How included and able to contribute do new and existing members feel?'') and \textit{Diversity} (i.e., ``How diverse are the people in the community?''). The response was recorded on an 11-point scale (from 1-11) for consistency with prior work~\cite{weld2022makes}. 

\subsubsection{Community Moderation}
We use content removal as our main measure of community moderation. Using Pushshift data from Jan. 2022 - Dec. 2023, we compute \textit{Content Removal Rate} as the percentage of submissions removed by moderators within a subreddit~\cite{jhaver2019does}. \new{In total, there were $856,980$ submissions across twelve subreddits. The removal rate ranged from $16.0\%$ (\texttt{r/conspiracy}) to $41.5\%$ (\texttt{r/changemyview}), with a median of $31.9\%$ percent per subreddit (see Appendix~\ref{subsec:app_descriptive}} for disaggregated removal rates). We apply a logarithmic transformation with base 2 after adding a start-value of 1 to the computed values.

\subsubsection{Controls}
Prior work on the spiral of silence has surfaced individual-level factors that influence a person's willingness to voice their viewpoint~\cite{dalisay2012spiral,salmon1990community,matthes2010spiral}. First, we include a measure for a participant's willingness to self-censor (\textit{WTSC}) to account for individual personality differences that could impact the likelihood of sharing. Following prior work~\cite{gearhart2014gay,sherrick2018effect,chen2018spiral}, we use the composite measure from Hayes et al.~\cite{hayes2005willingness}, which consists of eight questions that capture an individual's likelihood of withholding their opinion when they sense others disagree. A larger \textit{WTSC} value means the individual is more disposed to self-censoring. Second, following \citet{matthes2010spiral}'s finding that how strongly an individual agrees with a viewpoint impacts the chance that the spiral of silence occurs, we include a control for \textit{Agreement Intensity} measured on a three-point scale. Finally, we include controls for user demographics, as prior work~\cite{pierson2015outnumbered,de2017gender} has found differences in sharing behavior across identity groups. We include three binary variables for gender, race, and political leaning. 

We also include controls related to the participant's Reddit usage. Importantly, we expect that many participants are lurkers, who are unlikely to post regardless of the topic~\cite{nonnecke2000lurker,gong2015characterizing}. To control for the different base rates of posting frequency, we include a measure, \textit{Posting}, which captures a participant's self-reported likelihood of actively participate in a community (as opposed to lurking) on a 5-point scale. Finally, we expect newcomers may exhibit different sharing behavior compared to platform veterans~\cite{yang2017commitment}. Thus, we include the control \textit{Account Tenure} that measures the age of the participant's Reddit account in years.

\section{Results}
\label{sec:results}
\begin{figure*}[h]
    \centering
    \includegraphics[width=\textwidth]{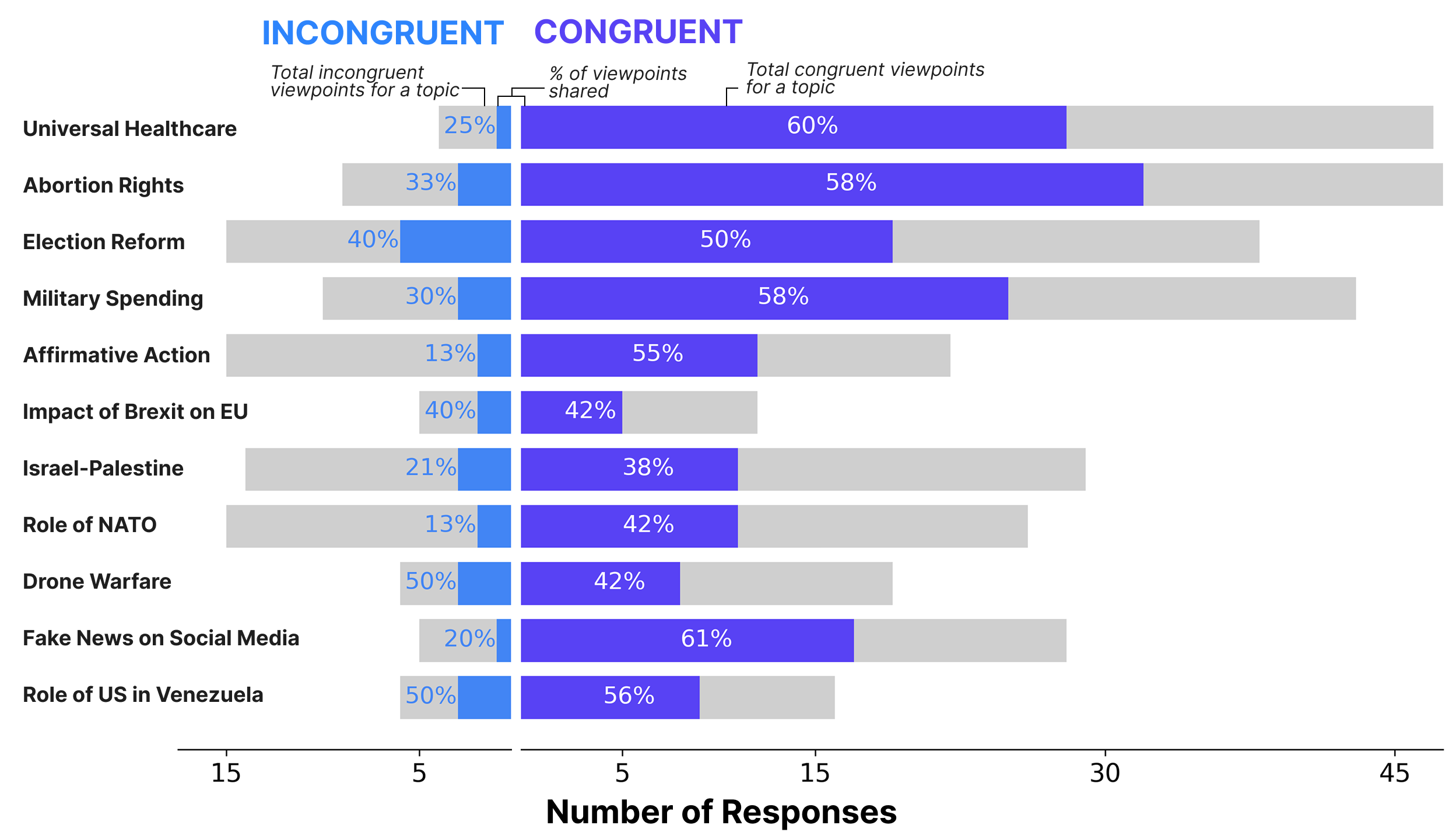}
    \caption{Across all topics, the average percentage of incongruent viewpoints likely to be shared is $27.9\%$. On the left, we report the percentage of incongruent viewpoints willing to be shared (i.e., \textit{Share Likelihood} $>$ 4) out of all incongruent viewpoints for the given topic. On the right, we do the same for congruent viewpoints.}
    \label{fig:viewpoint_breakdown}
    \Description[Bar chart comparing the likelihood of sharing congruent versus incongruent viewpoints across eleven topics.]{Bar chart comparing the likelihood of sharing congruent versus incongruent viewpoints across eleven topics. On average, 27.9\% of incongruent viewpoints are likely to be shared compared to a substantially higher proportion of congruent viewpoints.}
\end{figure*}

\new{From our survey, we find that participants are less likely to voice incongruent viewpoints across topics. Among responses where the reported viewpoint aligns with the majority opinion in a subreddit, 47.2\% indicate a high likelihood of sharing their viewpoint (Share Likelihood > 4). In contrast, only 27.9\% of responses expressing incongruent viewpoints report a high likelihood of sharing. Looking across topics, congruent viewpoints are $2.03\pm0.98$ times more likely to be shared compared to incongruent viewpoints. The difference in sharing is most stark for the topic of affirmative action, where congruent viewpoints are $4.09$ times more likely to be shared compared to incongruent viewpoints. We observe only one topic (``ethics of drone warfare'') in which participants are more likely to share incongruent viewpoints compared to congruent ones.} 

In this section, we formalize these analyses using linear mixed-effect models. We start by presenting the model we used for analysis. Next, we compare opinion expression in congruent versus incongruent opinion climates. Then, we explore the association of community-level factors, such as inclusivity and moderation practices, with self-silencing. We conclude by analyzing what factors are associated with variation in upvoting behavior.

\subsection{Method of Analysis}
To analyze the results of our survey, we use linear mixed-effect models, with subreddit and participant as crossed random effects. We introduce a series of three nested models. As shown in Table~\ref{tab:model1}, we first have a model using only our control variables as predictors. We then add additional measures for viewpoint incongruency, community inclusivity, diversity, and moderation. Finally, we include interaction effects between incongruency and our measurements for community values and moderation to our model. We standardize all continuous independent variables and controls. For our first set of models (Models 1a-1c), the dependent variable is the self-reported likelihood of sharing the viewpoint on a main account (\textit{Share Likelihood}). In Models 2a-2c, we use the likelihood of upvoting the viewpoint (\textit{Upvote Likelihood}) as our dependent variable, which we report in Table~\ref{tab:model2}. The dependent variables are left in raw units.

\begin{table*}[t!]
    \centering
    \caption{Although incongruent viewpoints are reported less often than congruent ones, communities with greater perceived diversity are associated with lower levels of self-silencing. We present the results of a linear mixed-effect model that predicts a participant's likelihood of sharing their opinion as a hierarchical regression. Model 1a contains only our controls. Model 1b includes the controls and fixed effects for opinion incongruency, community values, and community moderation. Model 1c adds interaction effects. Note: $^{*}$p$<$0.05; $^{**}$p$<$0.01; $^{***}$p$<$0.001.}
        \begin{tabular}{lrrlrrlrrl}
    \toprule
         & \multicolumn{3}{c}{\textbf{Model 1a}} &  \multicolumn{3}{c}{\textbf{Model 1b}} &  \multicolumn{3}{c}{\textbf{Model 1c}}\\
    \cmidrule(lr){2-4}\cmidrule(lr){5-7}\cmidrule(lr){8-10}
     & Coef. & SE & $p$ & Coef. & SE & $p$ & Coef. & SE & $p$\\ \midrule
    (Intercept) & $4.56$ & $0.60$ & $0.00^{***}$ & $4.78$ & $0.59$ & $0.00^{***}$ & $4.83$ & $0.59$ & $0.00^{***}$\\
    Account Tenure & $0.09$ & $0.24$ & $0.71$ & $0.09$ & $0.24$ & $0.72$ & $0.10$ & $0.24$ & $0.68$\\
    Posting & $0.70$ & $0.24$ & $0.01^{**}$ & $0.64$ & $0.24$ & $0.01^{**}$ & $0.64$ & $0.24$ & $0.01^{**}$\\
    Male & $-0.70$ & $0.56$ & $0.21$ & $-0.75$ & $0.55$ & $0.17$ & $-0.77$ & $0.54$ & $0.16$\\
    White & $0.11$ & $0.54$ & $0.83$ & $0.12$ & $0.53$ & $0.82$ & $0.12$ & $0.52$ & $0.82$\\
    Democrat & $-0.66$ & $0.50$ & $0.19$ & $-0.74$ & $0.50$ & $0.14$ & $-0.76$ & $0.49$ & $0.13$\\
    WTSC & $-0.27$ & $0.25$ & $0.28$ & $-0.30$ & $0.25$ & $0.22$ & $-0.33$ & $0.24$ & $0.19$\\
    Agreement & $0.48$ & $0.06$ & $0.00^{***}$ & $0.41$ & $0.06$ & $0.00^{***}$ & $0.42$ & $0.06$ & $0.00^{***}$\\ \addlinespace
    Incongruency &  &  &  & $-0.76$ & $0.14$ & $0.00^{***}$ & $-0.66$ & $0.14$ & $0.00^{***}$\\
    Inclusion &  &  &  & $0.09$ & $0.11$ & $0.42$ & $0.03$ & $0.11$ & $0.77$\\
    Diversity &  &  &  & $-0.02$ & $0.10$ & $0.85$ & $-0.08$ & $0.10$ & $0.42$\\
    Content Removal Rate &  &  &  & $-0.05$ & $0.08$ & $0.53$ & $0.04$ & $0.09$ & $0.69$\\ \addlinespace
    Incongruency $\times$ Inclusion &  &  &  &  &  &  & $0.23$ & $0.14$ & $0.12$\\
    Incongruency $\times$ Diversity &  &  &  &  &  &  & $0.29$ & $0.14$ & $0.03^{*}$\\
    Incongruency $\times$ Content Removal Rate &  &  &  &  &  &  & $-0.31$ & $0.14$ & $0.03^{*}$\\
    \midrule
    Marginal $R^2$ & $0.219$ &  &  & $0.244$ &  &  & $0.254$ &  & \\
    Conditional $R^2$ & $0.799$ &  &  & $0.810$ &  &  & $0.810$ 
    &  & \\
   
    \bottomrule
    \end{tabular}
    \label{tab:model1}
    \Description[Results of hierarchical linear mixed-effects models predicting likelihood of sharing a viewpoint.]{
    Results of hierarchical linear mixed-effects models predicting likelihood of sharing a viewpoint. The table reports coefficients, standard errors, and significance levels across three nested models, showing that incongruent viewpoints are significantly less likely to be shared. The interaction effect between community diversity and incongruency is significant and positive whereas the interaction between content removal rate and incongruency is significant and negative.
    }
\end{table*}

\subsection{Analyzing Opinion Sharing Behavior}
\label{subsec:sharing}
\subsubsection{Participants are willing to speak up for congruent viewpoints.} We start by examining the relationship between the likelihood of sharing a viewpoint and our control variables. In Model 1a, the intercept is $4.56$, representing the likelihood of sharing a congruent viewpoint for a non-Male, non-White, and non-Democrat individual who posts on Reddit rarely to occasionally, has an account age of $6.2$ years, and mean WTSC ($3.68$ out of 8). The intercept value falls between ``neither likely nor unlikely'' and ``slightly likely'' on the 7-point scale from the survey. As expected, posting frequency is positively associated with opinion expression ($\beta=0.70$, $p=0.01$). In addition, participants who agree with a viewpoint more strongly are also more likely to comment ($\beta=0.42$, $p<0.001$). However, outside of posting frequency and agreement intensity, we do not find any statistically significant differences for our control variables.  

\subsubsection{Incongruent viewpoints are shared less often.}
Participants were less likely to voice incongruent viewpoints across all topics despite agreeing with the stance. Approximately half ($52.8\%$) of participants responded that they are likely to share their viewpoint (\textit{Share Likelihood} $>4$) when they believe themselves to agree with the majority (see Fig.~\ref{fig:viewpoint_breakdown}). In contrast, the proportion of participants likely to share incongruent viewpoints is lower at $27.9\%$, although the percentage ranges across topics from $13.3\%$ to $50.0\%$. On average, across all topics, participants are $2.04$ times more likely to share a viewpoint they perceive as being in the majority compared to viewpoints they believe are in the minority.

This result is encoded in Model 1b, which includes a fixed effect for viewpoint incongruency, allowing us to compare the likelihood of sharing a viewpoint that is perceived to be in the majority versus minority. A likelihood ratio test (ANOVA) confirms that adding these fixed effect significantly improves the explanatory power from Model 1a, which only includes demographics and other control variables as predictors, to 1b  ($\chi^2=31.4$, $p < 0.001$). In line with the spiral of silence theory, we observe a negative relationship between opinion incongruency and the likelihood of sharing ($\beta=-0.76, p<0.001$), indicating that participants may be less inclined to express their viewpoint when they believe their opinion differs from that of the majority. On average, when participants perceive themselves as being in the minority, they are less likely to share their viewpoint ($M=3.01$) compared to those who view themselves as in line with the majority ($M=4.15$). 

While participants view themselves as being congruent with the majority for most viewpoints, $72.4\%$ (N=42) of participants hold at least one incongruent viewpoint. Given that users must self-select into being members of different communities on Reddit, it is unlikely that they will be completely incongruent across all topics. Although community members are not typically in the opinion minority, they are likely to feel incongruent on at least a small selection of topics discussed on the subreddit.

\begin{figure*}[t!]
    \centering
\includegraphics[width=0.95\linewidth]{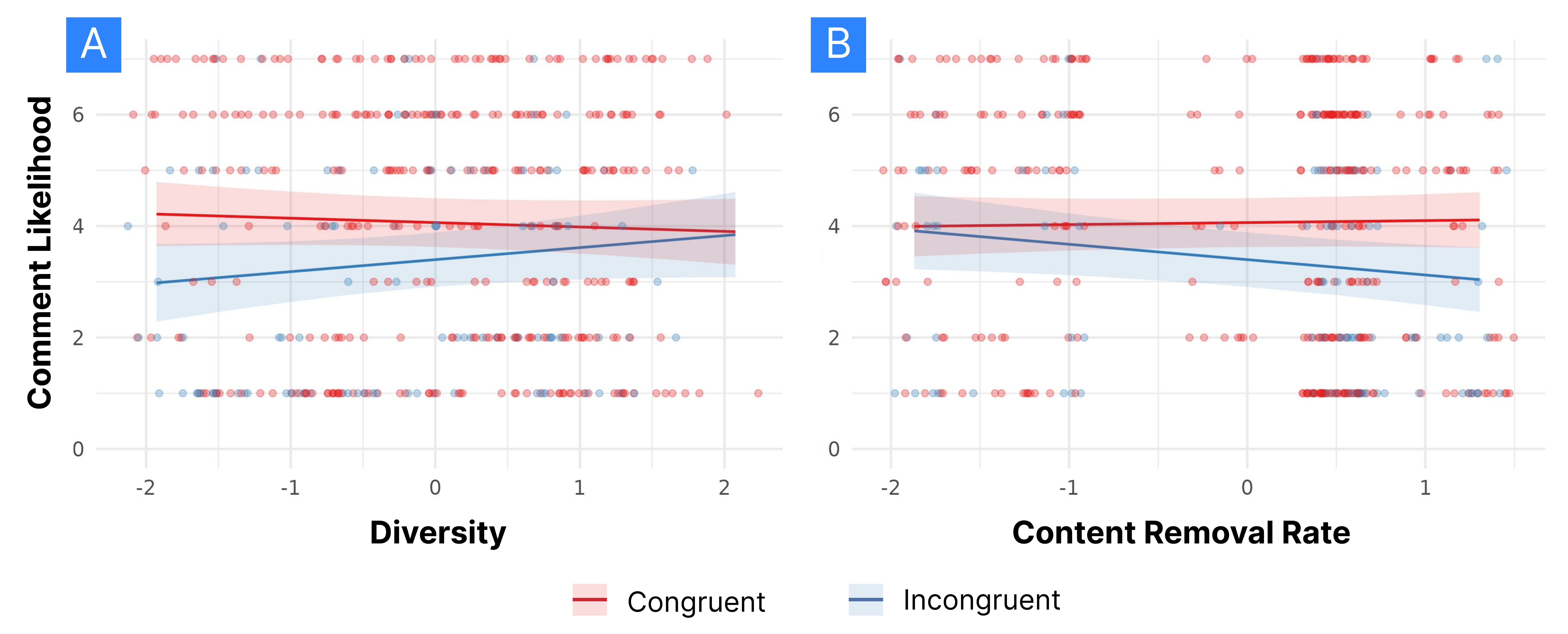}
    \caption{Community design factors relate to participants' likelihood of sharing incongruent viewpoints. The left panel (A) shows the marginal effects of community diversity on opinion expression for incongruent and congruent viewpoints. The right panel (B) shows the marginal effects of content removal rate on opinion expression for incongruent and congruent viewpoints.}
    \label{fig:marginal}
    \Description[Two-panel visualization showing the interaction effects between opinion incongruency and community design factors (diversity and content removal rate).]{Two-panel visualization showing the interaction effects between opinion incongruency and community design factors (diversity and content removal rate). Panel A plots the positive relationship between perceived community diversity and willingness to share incongruent viewpoints. Panel B shows that higher content removal rates are associated with a reduced likelihood of sharing incongruent viewpoints.}
\end{figure*}

\subsubsection{Community diversity is positively associated with incongruent sharing behavior.}
In addition to incongruency, we include measures of participants' perceived inclusivity and diversity of the subreddit. The main effects of perceived diversity ($\beta=-0.08$, $p=0.42$) and inclusivity ($\beta=0.03$, $p=0.77$) are not statistically significant. However, we find a significant interaction between diversity and incongruency ($\beta=0.29$, $p=0.03$), supporting H2b that higher perceived community diversity is associated with a greater likelihood of sharing incongruent viewpoints. These results come from Model 1c, which includes interaction effects between incongruency and our community design factors.\footnote{We confirm that adding these additional predictors significantly improves model fit compared to Model 1b ($\chi^2=13.3$, $p=0.004$).} Specifically, for incongruent viewpoints, a one-standard-deviation increase in perceived community diversity is associated with a \(0.29\)-point increase in the likelihood of commenting, whereas diversity shows a negligible relationship for congruent viewpoints ($\beta=-0.08$). This result suggests that when individuals perceive communities as being more heterogeneous, this perception is associated with a higher likelihood of sharing incongruent viewpoints.

\new{We do not find evidence at our level of statistical power to support a claim that more inclusive communities are associated with a greater likelihood of incongruent opinion-sharing. The interaction effect between \textit{Incongruency} and \textit{Inclusion} indicates a positive but non-significant relationship between perceived community inclusion and participants' willingness to share minoritized viewpoints ($\beta=0.23$, $p=0.12$). It is possible that in politics-oriented subreddits, community inclusion is less of a concern to users. When comparing the relative importance of different values across subreddits,~\citet{weld2021making} found that subreddits focused on news, media, and discussion, such as those included in our analysis, placed lower priority on inclusion as a community value.}

\subsubsection{Content removal has a negative relationship with incongruent viewpoint sharing behavior.}
\new{Finally, we look at our measure \emph{Content Removal Rate} which captures the amount of moderation within a given subreddit. While the main effect for content removal rate is not statistically significant ($\beta=0.04$, $p=0.69$), there is a significant interaction between removal rate and incongruency ($\beta=-0.31$, $p=0.03$). Thus, the association between moderation and participants' likelihood of commenting depends on how incongruent their viewpoint is. This result contradicts our initial hypothesis that moderation activity would bolster viewpoint sharing by promoting a more open environment. In practice, moderation may be associated with the opposite: seeing more comments being removed could exacerbate participants' fear of social isolation, or reflect more heavy-handed ideological moderation, correlating with more self-silencing. To pursue this question further, we explored whether there is a quadratic relationship between moderation and participation, since draconian moderation policies could stifle people's desire to speak up~\cite{kraut2012building} (see Appendix~\ref{sec:app_quad} for full results). While there may be weak evidence that such trend exists for incongruent viewpoints ($\beta= -0.32$), the relationship is non-significant ($p=0.07$) at our level of power. It is likely that the subreddits in our sample are not overly moderated to the point that participation is diminished or subreddit members are unaware of the extent of content removal within a community~\cite{jhaver2019did}.}

\subsection{Analyzing Upvoting Behavior}
\begin{figure*}[ht!]
    \centering
    \includegraphics[width=\textwidth]{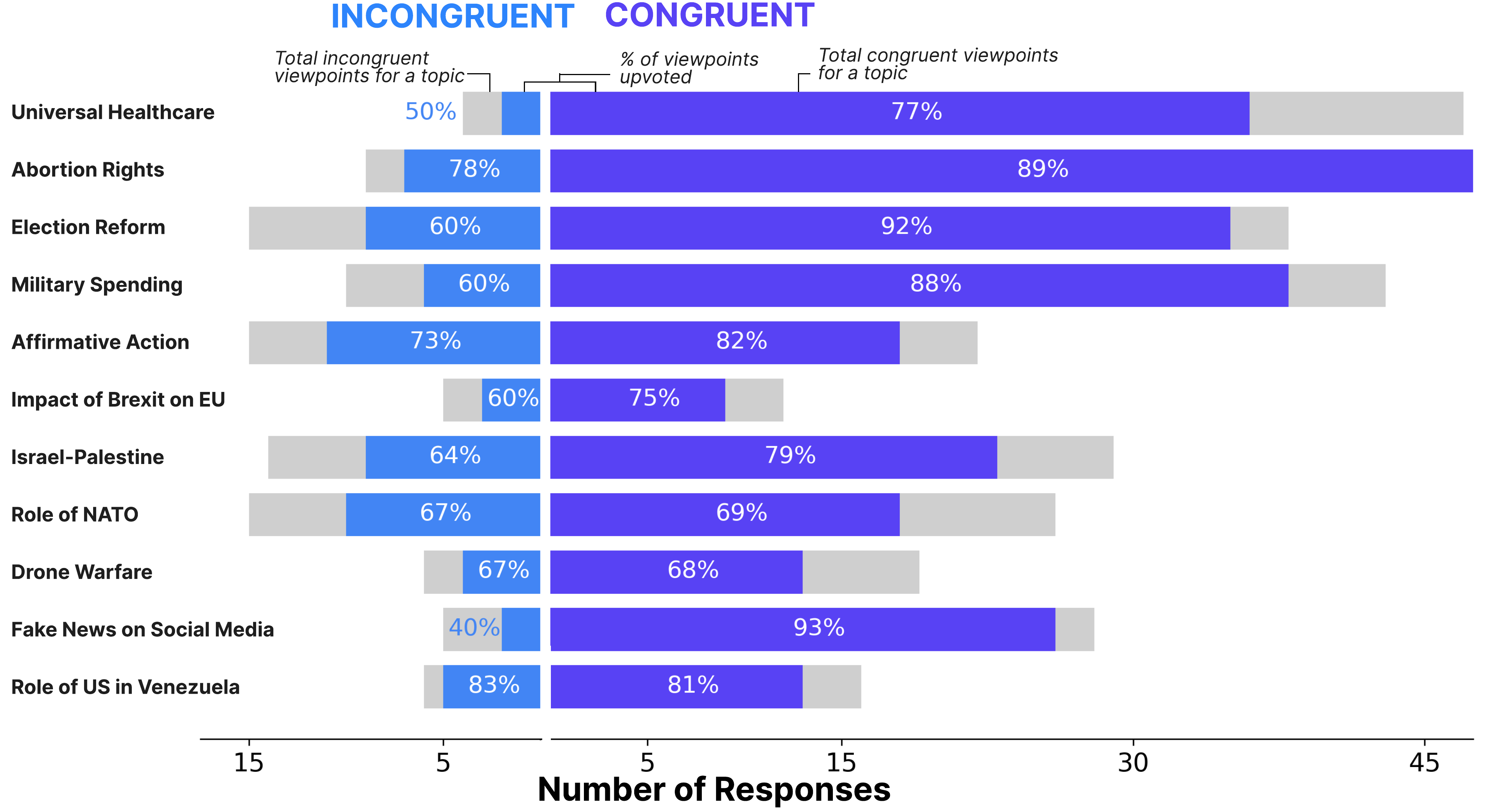}
    \caption{Across all topics, the average percentage of incongruent viewpoints likely to be upvoted is $65.4\%$. On the left, we report the percentage of incongruent viewpoints that participants are willing upvote (i.e., \textit{Upvote Likelihood} $>$ 4) out of all incongruent viewpoints for the given topic. On the right, we do the same for congruent viewpoints.}
    \label{fig:viewpoint_breakdown}
    \Description[Bar chart comparing upvoting behavior for congruent and incongruent viewpoints across eleven topics.]{Bar chart comparing upvoting behavior for congruent and incongruent viewpoints across eleven topics. Overall, congruent viewpoints are more likely to be upvoted (83.0\%) than incongruent ones (65.4\%).}
\end{figure*}
\begin{table*}[h!]
    \centering
    \caption{Incongruent viewpoints are less likely to be upvoted compared to congruent viewpoints. We present the results of a linear mixed-effect model that predicts the likelihood of upvoting as a hierarchical regression. Model 2a only includes the controls; Model 2b has the controls and fixed effects for \textit{Incongruency}, community values, and community moderation; and Model 2c adds interaction effects. Note: $^{*}$p$<$0.05; $^{**}$p$<$0.01; $^{***}$p$<$0.001}
    \begin{tabular}{lrrlrrlrrl}
    \toprule
         & \multicolumn{3}{c}{\textbf{Model 2a}} &  \multicolumn{3}{c}{\textbf{Model 2b}} &  \multicolumn{3}{c}{\textbf{Model 2c}}\\
        \cmidrule(lr){2-4}\cmidrule(lr){5-7}\cmidrule(lr){8-10}
     & Coef. & SE & $p$ & Coef. & SE & $p$ & Coef. & SE & $p$\\ \midrule
    (Intercept) & $5.63$ & $0.43$ & $0.00^{***}$ & $5.72$ & $0.42$ & $0.00^{***}$ & $5.73$ & $0.42$ & $0.00^{***}$\\ 
    \textbf{Controls} \\
    WTSC & $0.07$ & $0.18$ & $0.68$ & $0.07$ & $0.18$ & $0.69$ & $0.07$ & $0.18$ & $0.70$\\ 
    Agreement Intensity & $0.48$ & $0.05$ & $0.00^{***}$ & $0.46$ & $0.05$ & $0.00^{***}$ & $0.46$ & $0.05$ & $0.00^{***}$\\
    Male (0/1) & $-0.12$ & $0.40$ & $0.76$ & $-0.15$ & $0.39$ & $0.71$ & $-0.15$ & $0.39$ & $0.71$\\
    White (0/1) & $-0.29$ & $0.39$ & $0.46$ & $-0.27$ & $0.38$ & $0.48$ & $-0.27$ & $0.38$ & $0.48$\\ 
    Democrat (0/1) & $0.31$ & $0.36$ & $0.38$ & $0.24$ & $0.35$ & $0.51$ & $-0.23$ & $0.35$ & $0.52$\\
    Account Tenure & $-0.02$ & $0.17$ & $0.89$ & $-0.03$ & $0.17$ & $0.88$ & $-0.03$ & $0.17$ & $0.88$\\
    Posting & $0.57$ & $0.17$ & $0.00^{**}$ & $0.55$ & $0.17$ & $0.00^{**}$ & $0.55$ & $0.17$ & $0.00^{**}$\\ \addlinespace
    \textbf{Independent Variables} \\
    Incongruency (0/1) & & & & $-0.22$ & $0.12$ & $0.07$ & $-0.20$ & $0.12$ & $0.10$\\
    Inclusion & & & & $0.09$ & $0.08$ & $0.27$ & $-0.08$ & $0.08$ & $0.32$\\
    Diversity & & & & $-0.00$ & $0.07$ & $0.99$ & $-0.02$ & $0.08$ & $0.83$\\
    Content Removal Rate & & & & $0.06$ & $0.06$ & $0.37$ & $0.08$ & $0.07$ & $0.28$\\ \addlinespace
    Incongruency $\times$ Inclusion  & & & & & & & $0.02$ & $0.12$ & $0.89$\\
    Incongruency $\times$ Diversity & & & & & & & $0.08$ & $0.12$ & $0.50$\\
    Incongruency $\times$ Content Removal Rate & & & & & & & $-0.08$ & $0.12$ & $0.50$\\
    \midrule
    Marginal $R^2$ & $0.216$ & & & $0.226$ & & & $0.227$\\
    Conditional $R^2$ & $0.736$ & & & $0.735$ & & & $0.733$\\
    \bottomrule
    \end{tabular}
    \label{tab:model2}
    \Description[Results of hierarchical linear mixed-effects models predicting likelihood of upvoting a viewpoint.]{
    Results of hierarchical linear mixed-effects models predicting likelihood of upvoting a viewpoint. The table compares three nested models and shows that incongruent viewpoints are somewhat less likely to be upvoted, though the effect is weaker and not statistically significant.
    }
\end{table*}
So far, we have focused on posting as the primary way users express their opinions, but participants can signal their viewpoints in other ways. We also examine upvoting as an alternative form of opinion expression. Unlike posting or commenting, upvoting is anonymous. From our survey, we find that participants are likely to upvote $65.4\%$ of incongruent viewpoints, compared to $83.0\%$ of congruent viewpoints. Consistent with our findings on sharing (Sec.~\ref{subsec:sharing}), when disaggregated by topics, congruent viewpoints are still more likely to be upvoted. However, this difference is smaller, with congruent viewpoints only $1.33\pm0.37$ times more likely to receive an upvote.

\new{To study the relationship between upvoting and our selected predictors, we again use a set of nested linear mixed-effects model with subreddit and participant as crossed random effects. Our dependent variable is the likelihood of upvoting variable (measured on a 7 point Likert scale). In our models, we use the same control variables and predictors from our previous analysis on sharing likelihood (see Sec.~\ref{subsec:sharing}). The following analysis was not pre-registered and should be considered post-hoc.} 

\new{\subsubsection{Upvoting is more common than commenting} We begin by examining how likely participants are to upvote comments across subreddits. As seen in Table~\ref{tab:model2}, the intercept is $5.63$ for Model 2a. This indicates that it is ``slightly'' to ``moderately'' likely that a non-Male, non-White, and non-Democrat individual who posts on Reddit rarely to occasionally, has an account age of 6.2 years, and WTSC of 3.68 out of 8 would upvote a congruent viewpoint. Similar to our results in Sec.~\ref{subsec:sharing}, we find a significant positive correlation between posting frequency and upvoting ($\beta=0.57$, $p < 0.001$) as well as agreement intensity and upvoting ($\beta=0.48$, $p < 0.001$). Overall, participants are more likely to upvote a comment ($M=5.47$) compared to posting the comment on their main account ($M=3.87$).} 

\subsubsection{Incongruent viewpoints are less likely to be upvoted, even when people agree with the stance.} \new{Similar to participants' behavior with sharing viewpoints on Reddit, we observe that participants are less likely to upvote incongruent comments ($M=4.93$) compared to congruent ones ($M=5.63$). While we also observe a negative relationship between \textit{Incongruency} and the likelihood of upvoting ($\beta=-0.22$ in Model 2b), this association is not statistically significant, likely due to lack of statistical power ($p=0.07$). Furthermore, unlike with commenting, we do not observe any interaction effects between our community design factors (i.e., inclusion, diversity, and moderation) and opinion congruency. This result indicates that these factors are not associated with participants' decision to upvote a congruent viewpoint compared to an incongruent one. Following Noelle-Neumann's theory~\cite{noelle1974spiral}, we do expect the spiral of silence to be less pronounced for upvoting compared to posting or commenting as it is completely anonymous, making it less likely that participants anticipate social isolation or negative reactions as a consequence of their actions.}

\subsubsection{Upvoting provides an alternative for sharing otherwise self-silenced opinions.} \new{Overall, upvoting provides a valuable avenue for users to express their opinion. We compare the number of viewpoints that users report that they are not willing to comment on but are willing to upvote. In congruent conditions, $65.8\%$ of viewpoints that would not be posted (\textit{Share Likelihood} $ \leq 4$) would be upvoted (\textit{Upvote Likelihood} $ > 4$). While this percentage of viewpoints is lower in comparison for incongruent viewpoints, for half of the incongruent viewpoints ($53.3\%$), participants are still likely to use upvoting as a mechanism for expressing their opinions, even when they are unwilling to post them.}

\section{Discussion}
In our work, we seek to measure the extent of self-silencing behavior across different political subreddits. Through our analysis, we find evidence that confirms the spiral of silence theory and uncovers community design factors that can help counteract these silencing effects on Reddit. A summary of our hypotheses and findings can be found in Table~\ref{tab:summary}. In this section, we cover both the theoretical implications for understanding the spiral of silence on social media and complementary design recommendations. 

\begin{table*}[t]
    \centering
    \caption{Summary of hypotheses and findings. $\checkmark$  
   indicates the hypothesis is supported and ``NS'' indicates the hypothesis is not supported.}
    \begin{tabular}{>{\raggedright}p{4in}>{\raggedright\arraybackslash}p{0.5in}}
    \toprule
    \textbf{Hypothesis} & \textbf{Result} \\
    \midrule
    \RaggedRight{\textbf{H1}: The likelihood of opinion expression is negatively associated with viewpoint incongruency.} & \large{\checkmark} \\
    \addlinespace
    \RaggedRight{\textbf{H2a}: The perceived inclusivity of a community is more positively associated with the likelihood of opinion expression when users hold incongruent viewpoints.} & \large{NS} \\
    \addlinespace
    \textbf{H2b}: The diversity of community members is more positively associated with the likelihood of opinion expression when users hold incongruent viewpoints. & \large{\checkmark}  \\
    \addlinespace
    \textbf{H3}: Moderation activity is more positively associated with the likelihood of opinion expression when users hold incongruent viewpoints. & NS \\
    \bottomrule
    \end{tabular}
    \label{tab:summary}
    \Description[Summary of hypotheses and findings. ]{
    Summary of hypotheses and findings. The table indicates which hypotheses are supported or not supported based on study results, showing that community diversity significantly increases willingness to share incongruent viewpoints. Hypotheses around perceived community inclusion and content removal rate are not supported.
    }
\end{table*}

\subsection{Community Factors and the Spiral of Silence}
\label{subsec:community_design}
We identify community design factors on Reddit that can inform design decisions to counteract silencing effects. Previous works~\cite{neubaum2022s,wu2018comment,fox2021fear,woong2011selective} have focused on how platform-level designs influences the spiral of silence. For example, \citet{neubaum2022s} studied how perceptions of message persistence decrease the likelihood that users will share incongruent viewpoints. In practice, it is difficult for social media users and moderators to change platform design. Intervening at the community design level provides a more actionable alternative. From our analysis, we found that perceived diversity is positively associated with the likelihood of sharing when individuals consider their opinion to be in the minority, whereas perceived content removal rates are negatively associated with sharing incongruent viewpoints. Since moderators have more power to influence the values and norms of online communities, this finding suggest that there are feasible interventions to mitigate self-silencing within a community without having to change platform-level design. 

\new{In line with Hypothesis 2b, we observe a positive correlation between participants' perceptions of community diversity (i.e., how diverse they believe people in the subreddit are) and their willingness to post incongruous viewpoints on politically-oriented subreddits. While this result suggests there are actionable changes to community design factors that can reduce self-silencing, we also acknowledge that fostering a hetereogenous subreddit poses a challenging task for moderators. As prior work has found, there are increasing amounts of fragmentation on the platform with smaller, more thematically specified subreddits forming and pulling away users from larger, more aggregated communities~\cite{singer2014evolution}. On one hand, this diversification means that users are likely to find a community where their viewpoints are not minoritized. However, this behavior may lead to more homogeneity within subreddits, discouraging community members from sharing viewpoints they perceive as being in the minority. One path forward, however, can be to encourage new users to join and participate within a community. As \citet{duguay2022read} found, long-term active users within a subreddit express express more opinion homogeneity whereas heterogeneity within communities can be linked to new users joining. Thus, moderators can employ design interventions that encourage new joiners within the community~\cite{kraut2012building}. Examples that have been helpful on Reddit and online peer-production communities include having beginner FAQs~\cite{weld2021making}, introducing automated recommendations for topics new members may be interested in~\cite{yazdanian2019eliciting}, and using flairs / badges for newcomers~\cite{santos2020can}. Another lightweight intervention is to highlight diversity through positive reinforcement. Prior work~\cite{lambert2024proactively,lambert2025does} found that positive feedback on Reddit, such as awarding gold, can help set community norms and encourage participation --- a practice that especially benefits newcomers. A similar mechanism could be implemented by moderators to showcase diverse viewpoints in the community, helping combat the spiral of silence by demonstrating that this type of opinion-sharing is not only invited but also actively encouraged.} 

\new{Contrary to Hypothesis 3, we find that higher content removal rates within subreddits are associated with a decrease in participants' willingness to comment incongruent viewpoints. We initially expected that more content removal would assuage users' fear of retaliation or social isolation when sharing unpopular opinions. However, one explanation for this negative relationship is that users in subreddits with stricter moderation may be worried that their own comments will be removed by moderators, and this negative reaction is associated with a lower likelihood of speaking out~\cite{gearhart2015something}. A known issue with content removal is that users often find the process to be opaque~\cite{jhaver2019does,juneja2020through}. This lack of transparency leads users' to develop their own folk theories as to why their content was removed, often attributing it to them posting unpopular opinions or the moderators' own political biases~\cite{jhaver2019does}; in practice, it may be that the posts were violating community rules. Increasing transparency, even in the form of lightweight explanations as to why the removal occurred, can help users understand moderators' decisions. Otherwise, users may believe that expressing minoritized viewpoints will engender social sanctions, leading them to remain silent.}

\subsection{Platform Affordances and the Spiral of Silence}
Prior work has compared how affordances including network association, social presence, and message persistence influence the degree of self-silencing across social media platforms, including Facebook and Twitter~\cite{neubaum2022s,stoycheff2016under,oz2024platform}. Reddit's distinct affordances offer further insight into two platform characteristics that impact the spiral of silence. First, since Reddit consists of multiple online communities, the porous boundaries offer users more flexibility in choosing spaces where they feel their opinions are more in line with the majority. \new{Second, Reddit offers users a greater degree of perceived anonymity in comparison to other social media platforms, such as Facebook, where the spiral of silence has been previously studied~\cite{boulianne2024social,hoffmann2017spiral,gearhart2015something}.} Within Reddit, users are also afforded varying degrees of anonymity depending on their choice of commenting or upvoting. We build on prior work~\cite{fox2021fear,wu2018comment,woong2011selective}, which has explored how perceived anonymity influences self-silencing behaviors.

\subsubsection{Users have options to express their opinions across multiple communities} Belonging to multiple communities on one platform give users more flexibility to find a space where they may feel comfortable expressing their opinion. On Reddit, users can easily join and navigate across many subreddits, and, in fact, $93.1\%$ (N=54) of surveyed participants reported being active members of more than one community. This functionality is important to take into account because whether a viewpoint is incongruent varies depending on the community. For example, we find that a pro-life viewpoint is in the minority for many of the subreddits we examine (e.g., \texttt{r/politics}, \texttt{r/news}, etc.), but it is the majority for \texttt{r/conservative}. Albeit rare, we also found instances where participants report being more willing to share the same viewpoint in a different subreddit where their stance is aligned with the majority versus when they view themselves as being in the minor. Of note, we also found cases of the opposite --- where participants were more likely to share their viewpoint to a subreddit where it would be in the minority. We only observed this behavior for the subreddits \texttt{r/changemyview} and \texttt{r/conspiracy} where it is possible that sharing minoritized viewpoints may be more accepted. Overall, this malleability can be beneficial since it means users are likely to find outlets where they feel comfortable freely sharing their opinions. Alternatively, there is a concern that this behavior can contribute to an ``echo chamber''~\cite{cinelli2021echo} effect, as users only participate in communities where they know others will agree with them. 

\subsubsection{Self-silencing persists under pseudoanonymity} As prior work has pointed out~\cite{matthes2014methodological,fox2021fear}, how identifiable participants perceive themselves to be has an influence on their willingness to share their opinion. In contrast to other social media platforms, such as Facebook, where the spiral of silence has been studied before~\cite{hoffmann2017spiral,gearhart2015something}, Reddit provides a greater sense of perceived anonymity to users~\cite{boulianne2024social}. Yet, even under these conditions, our results indicate that participants are still less willing to align themselves with incongruent opinions. While Reddit accounts are not directly linked to users' offline identities, users may not want certain content linked to their main account or long-term identity on the platform~\cite{leavitt2015throwaway}. This rationale explains why users may choose to make ``throwaway accounts'' --- a temporary identity created under a different pseudonym --- on Reddit, especially when discussing stigmatized topics~\cite{ammari2019self,de2014mental}. In fact, when we surveyed participants what other actions they would take if they were not willing to comment an incongruent viewpoint under their main account, participants selected posting from a throwaway account as the preferred course of action for eleven viewpoints. The persistence of content on social media --- in comparison to offline communication where there is no permanent trace of what people have said --- can exacerbate self-silencing behaviors, especially on platforms with higher degrees of association, as users may worry that people they know will find or look through prior posts~\cite{fox2021fear,stoycheff2016under}.

We also observe that participants are less likely to upvote incongruent opinions compared to congruent ones. Given that upvoting is completely anonymous, fear of social isolation is likely to be less pressing of a concern for users. This finding is also consistent with results from \citet{woong2011selective} who found that participants were unwilling to share opinions they perceived to be in the minority even in a completely anonymous online forum, suggesting that despite platform affordances, people may feel a fear of isolation which leads to self-silencing. Another explanation for this behavior may be that upvoting does not solely indicate agreement but also serves as an indication that the content is worth sharing in the community~\cite{moore2017redditors}. Since participants believe incongruent viewpoints are only held by a minority of subreddit members, they might also view the content as less worthwhile to share, decreasing the likelihood of upvoting.

\subsection{Confirming and Measuring the Spiral of Silence Online} 
We find robust evidence that participants are less likely to share their viewpoint when they perceive their opinion to be in the minority. Corroborating~\citet{gearhart2018same}'s findings, we observe that the rate of self-silencing differs across issue types, although as a whole participants are consistently less willing to express incongruent viewpoints. The impact of the spiral of silence also differs across communities. For example, on \texttt{r/conservative}, the mean likelihood of sharing a congruent opinion is $3.98 \pm 0.30$ versus $3.10 \pm 0.64$ for an incongruent opinion. However, for \texttt{r/politics}, the gap is much larger between sharing a congruent ($3.15 \pm 0.24$) versus incongruent viewpoint ($1.68 \pm 0.42$). Overall, there is a consistent pattern that participants are less willing to express incongruent viewpoints.

We also introduce a method for identifying controversial viewpoints within a community and then measuring the extent of the spiral of silence. This method provides a flexible tool for generating topics that can be tailored to a specified community. In this work, we focused on identifying political and social issues, but our method can be applied to other topic areas. For example, we can generate plausible viewpoints even when using communities that focus on video games and sports --- \texttt{r/gaming} and \texttt{r/nba}), respectively (see Appendix~\ref{sec:hai_design} for details). Future work can adapt this method to explore other online communities and topics. 

\subsection{When (or When Not) to Reduce Self-Silencing}
As stated in Sec.~\ref{sec:methods}, the scope of our work is intentionally limited to viewpoints that plausibly fall within the bounds of a community while still eliciting meaningful disagreement. We provide design interventions in Sec.\ref{subsec:community_design} to mitigate the spiral of silence for such opinions, as suppressing these viewpoints can distort broader understandings of public opinion and prevent communities from engaging with the full spectrum of legitimate viewpoints. However, it is important to acknowledge that reducing self-silencing is not universally desirable. Encouraging the expression of harmful content, such as hate speech or misinformation, or viewpoints that violate community norms can be counterproductive, as toxicity often amplifies rather than reduces self-silencing among other users~\cite{olson2017feminazis,gearhart2015something}. Prior work has shown that elevated toxicity in online political conversations leads many to withhold participation to avoid harassment or conflict~\cite{mamakos2023social,juncosa2024toxic}. Moreover, the spiral of silence may operate differently for harmful content. For example, \citet{chaudhry2020expressing} found that Facebook users remained willing to express racist discourse, despite it constituting a minority opinion, suggesting that perceived social sanctions do not uniformly suppress all types of opinion expression. These findings underscore an important caveat: interventions to reduce self-silencing must carefully distinguish between encouraging legitimate minority viewpoints be heard versus moderating harmful content within a community.
\subsection{Limitations and Future Work}
\subsubsection{Use of Self-Reported Measures}
\new{Our work provides a generalizable method for generating controversial topics and then measuring rates of self-silencing in communities using survey measures.} Following prior work~\cite{gearhart2014gay,gearhart2018same,chia2014authoritarian}, we use self-reported measures of opinion expression for sharing and upvoting to capture the spiral of silence. This approach may lack ecological validity compared to studying sharing behavior directly on the platform, as participants' reported willingness to share may differ from their situated behaviors. Prior work~\cite{scheufele2001real} has also critiqued the use of a hypothetical scenario when measuring likelihood of opinion expression as it may not faithfully capture how participants would act in a realistic setting. Furthermore, by only studying self-reports, we are unable to analyze what participants would say if they were to express their opinion; our current measure is unable to capture the nuances between participants boldly stating their stance on a topic versus making a hedging or lukewarm statement. Nonetheless, our approach remains well validated in prior work~\cite{gearhart2014gay,gearhart2018same,chia2014authoritarian}. 

A fruitful avenue for future work is to develop new methods for measuring participants' opinion expression behavior beyond relying on hypothetical self-reports. For example, building systems that can directly measure self-silencing on the social media platform would provide more realistic accounts of user behavior compared to surveys. Another option could be to collect observed behavior traces (e.g., past comments and upvotes on Reddit) or account information (e.g., degree of anonymity on their profile) that can be linked to self-report data. This information allows us to better model users' behavior on the platform,  providing a more holistic picture about the users' self-silencing behaviors.

\subsubsection{Study Methodology}
In addition, there are limitations with our methodology that can be explored in future studies. First, we take an automated approach to generating topics and viewpoints. While we include multiple validation steps to ensure that these topics are relevant and plausible for a given subreddit, this method prioritizes precision over recall; we are not guaranteed to capture the full range of controversial topics within a community. For example, there may be niche topics specific to certain online communities that our current approach misses. Second, we recruited our survey participants using the Prolific platform, which can introduce selection bias. This recruitment method also limited the amount of information we could collect about participants (e.g., Reddit username), meaning that although we screened participants using our knowledge quiz to validate subreddit activity, we are still relying on self-report data. Third, in our survey, we ask participants to select from two opposing viewpoints, which may not fully capture the spectrum of stances people have on a topic. We note that in empirical studies of the spiral of silence, it is common to operationalize a topic into a binary choice for survey respondents~\cite{chia2014authoritarian,gearhart2018same,matthes2010spiral,neuwirth2007spiral}. Expanding the range of viewpoints considered could offer a more nuanced understanding of how different minority perspectives interact within online discussions or shed insight into whether the spiral of silence manifests when there is not a prevailing perceived majority opinion on a controversial topic. Our work also only provides insights on correlations between incongruency and opinion sharing; we are not able to make causal claims about this relationship. 

There are many extensions to the study methodology that would provide a richer understanding of the spiral of silence. For example, future work can explore how different operationalizations of measurements in our study, such as treating incongruency as continuous or ordinal rather than binary, or having different definitions of community design factors are related to self-silencing behavior. We can also consider new methods for obtaining surveyed topics; one possible extension is to augment the generated topics with crowdsourced viewpoints from community members, although particular care must be taken to ensure the responses themselves are not skewed by the silencing effects already occurring within the community. Finally, to better understand causality, future work can draw on experimental methods to test whether placing users in incongruent versus congruent environment shapes their likelihood of sharing opinion in a controlled fashion~\cite{zerback2017can,wu2018comment}. 

\subsubsection{Generalizability of Results}
Finally, it is important to note that our analyses are centered on politically-oriented communities on Reddit. This decision raises two limitations. First, we consider a limited subset of subreddits on Reddit. We chose to study politically-oriented communities, as \citet{noelle1974spiral} asserts that the spiral of silence occurs for morally-laden topics, such as those related to politics or social issues. Focusing on these communities allows us to study contexts where individuals are more likely to experience pressure to self-censor, making the phenomenon more observable. However, given the diversity of community values and norms on Reddit~\cite{weld2022makes}, this focus limits the extent to which our findings may apply to subreddits centered on neutral or less value-laden topics. Second, we focus our study on a single platform, Reddit, to explore how community design factors are associated with the spiral of silence. However, prior work has shown how the spiral of silence differs across social media platforms (e.g., Facebook compared to Twitter~\cite{oz2024platform}). Since platform affordances can mediate how the spiral of silence manifests, this may limit the transferability of the results from this work to other platforms. Overall, future exploration on other topic areas, communities, and platforms can help bolster the generalizability of these findings.

\section{Conclusion}
In this work, our goal is to measure the extent of the spiral of silence across different online political communities. We also seek to identify community design factors that may amplify or mitigate self-silencing effects. Since directly measuring what is left unsaid using social media data is not possible, we propose a new method using LLMs to propose community-specific topics and viewpoints that are likely to be controversial. Then, we survey users' likelihood of opinion expression, opinion incongruency, and perceived community values. In total, we collected 439 responses capturing subreddit community members' likelihood of opinion expression across twelve subreddits and eleven topics. We find robust evidence that participants are less likely to share their viewpoint when they consider themselves to be in the minority. This finding corroborates that the spiral of silence manifests even on Reddit, which affords users a greater degree of perceived anonymity compared to other social media platforms.

Although self-silencing behavior is prevalent across the subreddits and topics covered in our study, we identify community-level design decisions that can help mitigate the spiral of silence. We find that perceived diversity is positively associated with sharing minority opinions, while higher content removal rates are related to a decrease in willingness to share. Moderators can foster more heterogeneous communities by encouraging newcomer participation, using positive reinforcement to highlight diverse viewpoints, and increasing transparency in content removal decisions to help users understand that moderation targets rule violations rather than unpopular opinions. These design interventions can help a foster an environment where users do not feel as if they will be penalized for speaking out, provide actionable levers for encouraging opinion-sharing.

\begin{acks}
This work was funded by the Brown Institute for Media Innovation and Stanford HAI Seed Grant. Dora Zhao is also supported in part by the Paul and Daisy Soros Fellowship for New Americans. We thank Jordan Troutman, Tiziano Piccardi, Jacy Anthis, Lindsay Popowski, Omar Shaikh, and other members of Stanford HCI for their helpful comments and suggestions. We also thank Galen Weld for providing access to annotations from Media Bias/Fact Check. 
\end{acks}

\bibliographystyle{ACM-Reference-Format} 
\bibliography{egbib}
\appendix
\section{Additional Method Details}
\label{sec:app_details}
\subsection{Comparing Political Leanings}
\label{sec:teaser}
We compare the political leanings of active \texttt{r/politics} members with the orientation of content posted to the subreddit. To measure the political leanings of active members, we released a survey to Prolific users. To qualify as an ``active member'' of \texttt{r/politics}, participants must first self-identify as an active member of the subreddit and pass our knowledge quiz. For those participants that qualify, we asked for their self-reported political affiliation and the valency of this affiliation. In total, we surveyed 550 Prolific users, with 218 passing our screening requirements. 

To find the political orientation of content on r/politics, we leverage the fact that the subreddit’s rules require each submission consists of the headline from an article with the corresponding URL. Following prior works~\cite{weld2021political,cinelli2021echo}, we use partisan ratings from Media Bias/Fact Check (MBFC) to label the URLs. In total, we analyze 50,000 posts from Jan. 2022 - Dec. 2023. 59.1\% (N=29,589) of the URLs have corresponding MBFC ratings.

\subsection{Generating Controversial Viewpoints}
\label{sec:hai_design}
\subsubsection{Prompts}
\label{subsec:prompts}
Our first step is to generate a list of topics that are likely to lead to disagreement. We provide the subreddit name, description of the subreddit, and community rules retrieved using \texttt{PRAW}. For our study, we generate $20$ topics per subreddit using \texttt{GPT-4} with temperature set to 1. 
\newline
\begin{lstlisting}[style=gptprompt]
You are a member of r/${SUBREDDIT}. Your task is to generate controversial issues,
aka issues that will lead to disagreement between r/${SUBREDDIT} members.
<name>
${SUBREDDIT}
</name>
<description> 
${DESCRIPTION}
</description>
<rules>
${RULES}
</rules>
Provide ${NUMBER} issues that will be controversial in r/${SUBREDDIT}. 
Return just the issue, no justification.
\end{lstlisting}

We generate viewpoints representing different stances for each topic. Again, we provide the subreddit name with the description and community rules. We use the same set of hyperparameters from topic generation for this task. 
\newline
\begin{lstlisting}[style=gptprompt]
You are a member of r/${SUBREDDIT}. Your task is to generate viewpoints that
r/${SUBREDDIT} members would hold on a given topic.
<name>
${SUBREDDIT}
</name>
<description> 
${DESCRIPTION}
</description>
<rules>
${RULES}
</rules>
For each side of the issue, write a 50-word Reddit comment from the perspective of that side that 
follows the rules of r/${SUBREDDIT}. 
Issue: ${TOPIC}
Return only the comments as a list, no justification. Example output: ["I am pro-choice.", "I am pro-life."] 
\end{lstlisting}
 
Finally, we shorten each of the viewpoints that was generated using the following prompt. We use a temperature of $0$ for this task. 
\newline
\begin{lstlisting}[style=gptprompt]
Shorten a statement while retaining the same viewpoint. 
The statement should be phrased as an opinion.
Text: ${VIEWPOINT}
Return just the shortened statement, no justification.
\end{lstlisting}

\subsubsection{Generated Topics and Viewpoints}
\label{subsec:app_topiclist}
In Table~\ref{tab:viewpoint_list}, we provide the list of all eleven topics and their corresponding viewpoints. The first five topics were generated using \texttt{r/politics} as the seed subreddit and the subsequent six using \texttt{r/worldnews}. We also provide examples of applying our topic generation method to communities outside the realm of politics or social issues. In Table~\ref{tab:viewpoint_other}, we list five topics each from \texttt{r/nba} and \texttt{r/gaming}.

\begin{table*}[ht!]
    \centering
    \footnotesize
    \caption{The eleven topics, and their respective viewpoints, used in the opinion expression survey}
    \begin{tabular}{p{2in}p{3.5in}}
    \toprule
    \RaggedRight{\textbf{Topic}} &  \RaggedRight{\textbf{Viewpoint}} \\
    \midrule
    \multirow{2}{*}{Universal Healthcare} 
        & 1. I believe in Universal Healthcare because everyone deserves access to good health, funded by the government. \\[0.5em]
        & 2. I believe market competition and individual insurance plans are superior to Universal Healthcare. \\
    \addlinespace
    \multirow{2}{*}{Abortion Rights} 
        & 1. As a pro-choice supporter, I believe women's bodily autonomy and reproductive choices are crucial for gender equality. \\[0.5em]
        & 2. As a pro-life advocate, I believe every life from conception deserves legal protection. \\
    \addlinespace
    \multirow{2}{1.8in}{Election Reform and Voter ID Laws} 
        & 1. I believe in strict voter ID laws for a secure democracy and fewer fraud allegations. \\[0.5em]
        & 2. Stringent voter ID laws marginalize minority and low-income communities. We should make voting more accessible, not harder, and aim for higher turnout, not suppression. \\
    \addlinespace
    \multirow{2}{*}{Military Spending} 
        & 1. Increasing military spending is vital for national security, global presence, and economic growth. \\[0.5em]
        & 2. We should cut military spending to fund education, healthcare, and infrastructure. \\
    \addlinespace
    \multirow{2}{*}{Affirmative Action} 
        & 1. I believe affirmative action counters systemic biases and fosters a diverse, inclusive society. \\[0.5em]
        & 2. I believe affirmative action could unintentionally cause reverse discrimination and undermine merit, potentially increasing societal division. \\
    \addlinespace
    \multirow{2}{2in}{Impact of Brexit on the European Union} 
        & 1. Brexit, economically, appears detrimental to the EU, potentially signaling a decline in internationalism. \\[0.5em]
        & 2. Brexit could potentially boost EU cohesion as member states see the difficulties of exiting. \\
    \addlinespace
    \multirow{2}{2in}{Israeli-Palestinian conflict and the recognition of Jerusalem as Israel's capital} 
        & 1. Recognizing Jerusalem as Israel's capital disrupts the peace process by favoring Israel's contested claims over Palestinian rights. \\[0.5em]
        & 2. Recognizing Jerusalem as Israel's capital acknowledges Jewish ties to the city, but doesn't negate the need for fair negotiations. \\
    \addlinespace
    \multirow{2}{*}{Role of NATO in maintaining global peace} 
        & 1. NATO's collective defense and strategic alliances are crucial for global peacekeeping and international security. \\[0.5em]
        & 2. NATO's actions can sometimes escalate global tension by infringing on sovereignty and threatening peace. \\
    \addlinespace
    \multirow{2}{*}{Ethics of drone warfare in the Middle East} 
        & 1. Drone warfare is a necessary evil for global security due to its precision, efficiency, and safety for soldiers. \\[0.5em]
        & 2. Drone warfare inevitably causes collateral damage, violates human rights, and induces terror. \\
    \addlinespace
    \multirow{2}{2in}{Role of social media platforms in spreading fake news} 
        & 1. Social media's lax approach has led to a fake news epidemic, undermining informed decision-making and threatening societal stability. \\[0.5em]
        & 2. Blaming social media for fake news is misplaced; users should fact-check and stricter content control risks censorship. \\
    \addlinespace
    \multirow{2}{2in}{Role of the United States in the Venezuelan political crisis} 
        & 1. I believe US intervention is vital for Venezuela's democratic restoration and humanitarian aid. \\[0.5em]
        & 2. I view US involvement in Venezuela as neo-imperialism; each country should independently handle its internal affairs. \\
    \bottomrule
    \end{tabular}
    \Description[ Overview of the eleven political topics and their opposing viewpoints used in the opinion expression survey.]{
    Overview of the eleven political topics and their opposing viewpoints used in the opinion expression survey. The table lists eleven political topics used. Each is paired with two contrasting statements designed to capture different sides of the debate.
    }
    \label{tab:viewpoint_list}
\end{table*}
\begin{table*}[ht!]
    \centering
    \footnotesize
    \caption{Our viewpoint generation process can be used for non-political communities. We present a sample of topics, and their respective viewpoints, proposed for \texttt{r/NBA} and \texttt{r/gaming}.}
    \begin{tabular}{p{1.8in}p{4in}}
    \toprule
    \RaggedRight{\textbf{Topic}} & \RaggedRight{\textbf{Viewpoint}} \\
    \midrule
    \multicolumn{2}{l}{\textbf{r/NBA}} \\
    \midrule
    \multirow{2}{*}{LeBron James vs Michael Jordan} 
        & 1. LeBron's consistent dominance and achievements arguably make him the GOAT, surpassing MJ. \\[0.5em]
        & 2. Despite LeBron's feats, MJ's NBA Finals record, killer instinct, and game-changing ability make him the GOAT in my opinion. \\
    \addlinespace
    \multirow{2}{*}{NBA Season Length} 
        & 1. I believe the 82-game season should be shortened to maintain game quality and reduce player fatigue. \\[0.5em]
        & 2. The 82-game NBA season is crucial for testing team endurance, reducing fluke performances, and providing ample basketball for fans. \\
    \addlinespace
    \multirow{2}{1.8in}{International NBA games} 
        & 1. I believe international NBA games disrupt the season's rhythm, fatigue players with travel, and unfairly risk their health and team performance for sport's geographical expansion. \\[0.5em]
        & 2. I support international NBA games as they expand the audience, grow the brand, inspire youth, and boost the sport's global popularity, despite logistical challenges. \\
    \addlinespace
    \multirow{2}{*}{NBA Draft System} 
        & 1. I strongly favor abolishing the draft system for a free market system, allowing players to choose their teams and potentially balance league competition. \\[0.5em]
        & 2. The NBA draft is crucial for maintaining balance between small and large market teams and preventing the formation of super-teams. \\
    \addlinespace
    \multirow{2}{*}{Golden State Warriors 2015--2019} 
        & 1. The Warriors' 2015--2019 run boosted the league's competitiveness by forcing teams to adapt and improve. \\[0.5em]
        & 2. The Warriors' 2015--2019 dominance made the NBA predictable and potentially drove fans away. \\
    \midrule
    \multicolumn{2}{l}{\textbf{r/gaming}} \\
    \midrule
    \multirow{2}{*}{Console vs PC} 
        & 1. As a console gamer, I value its simplicity, exclusives, and plug-and-play convenience for relaxation. \\[0.5em]
        & 2. As a PC enthusiast, I believe PCs provide unmatched customization, performance, and game variety for in-depth gaming. \\
    \addlinespace
    \multirow{2}{2in}{Working conditions in gaming} 
        & 1. Crunch culture in game development is harmful and unsustainable; companies should prioritize employee wellbeing. \\[0.5em]
        & 2. Crunch may not be ideal, but without it, could top games be made? Many vocations are stressful, not just gaming. \\
    \addlinespace
    \multirow{2}{*}{Large vs indie game companies} 
        & 1. Major game companies' dominance stifles creativity; we need more space for indie developers to innovate. \\[0.5em]
        & 2. Big companies like Nintendo prove that dominance doesn't necessarily kill creativity, as they continually innovate and fund ambitious projects. \\
    \addlinespace
    \multirow{2}{*}{eSports as a real sport} 
        & 1. eSports, despite lacking physical exertion, is a sport due to its demand for strategy, teamwork, and skill. \\[0.5em]
        & 2. Esports should be classified as competitive gaming, not traditional sports due to the lack of physical exertion. \\
    \addlinespace
    \multirow{2}{*}{Objectification of female characters} 
        & 1. I believe game developers overly sexualize female characters, neglecting character depth. \\[0.5em]
        & 2. Video games are fantasy and idealized characters reflect artistic vision, not objectification. \\
    \addlinespace
    \multirow{2}{*}{Copyright issues and modding} 
        & 1. I support modders as they express themselves by reshaping games, while respecting authorship. \\[0.5em]
        & 2. I support developers because unauthorized mods can threaten their control over their costly game creations. \\
    \bottomrule
    \end{tabular}
    \label{tab:viewpoint_other}
    \Description[Examples of viewpoints for r/NBA and r/gaming]{
    The table lists examples of controversial viewpoints for two non-political communities -- r/Nba and r/gaming. 
    }
\end{table*}

\subsection{Identifying Relevant Subreddits}
\label{sec:app_subreddit_selection}
To find our list of political and social issues-oriented subreddits, we use the following process. Starting with 2,040 subreddits from \texttt{r/ListOfSubreddits}, we look at the 50 hottest submissions in each subreddit as of January 5, 2024. We select subreddits where the majority of submissions are in English -- classified using \texttt{langdetect}~\cite{nakatani2010langdetect} -- leaving us with 1,975 subreddits. Next, using \texttt{GPT-3.5-Turbo} we classify whether post titles are political (see below for prompt). We follow \citet{piccardi2024social} and use the definition of political content proposed by Pew Research Center~\cite{bestvater2022politics}. We choose subreddits where more than $80\%$ of submissions are labelled as being political (N=48). From the remaining subreddits, we select our final list of 23 based on subreddit size. In our final survey, participants also had the option to list an additional subreddit they were an active member of, bringing us to 33 subreddits in total. 

\begin{lstlisting}[style=gptprompt]
Political content on Reddit is varied and can be about officials and activists, social issues, or news and current events. Looking at the following post title, would you categorize it as POLITICAL or NOT POLITICAL content? 

Answer 1 if it is POLITICAL, 0 otherwise.

Post: ${POST}
Answer:
\end{lstlisting}

\subsection{Knowledge Quiz}
\label{sec:app_kq}
To validate that participants were members of the subreddits they selected, they were given a knowledge screening quiz. In the quiz, the participants were shown three post titles, two of which belonged to the top 50 hottest submissions of their selected subreddit and one that came from the top 50 hottest submissions of a different political subreddit. We manually checked that the title from the other subreddit was not relevant to the selected subreddit. The titles used in the knowledge quiz are listed in Table~\ref{tab:kq}.  
\begin{table*}[ht]
\footnotesize
    \centering
    \begin{tabular}{p{0.7in}p{1.5in}p{1.5in}p{1.5in}}
    \toprule
    \RaggedRight{\textbf{Subreddit}} & 
    \RaggedRight{\textbf{Title 1}} & 
    \RaggedRight{\textbf{Title 2}} & 
    \RaggedRight{\textbf{Title 3}}\\
    \midrule
    worldnews	& Mongolia Commits to Fighting Corruption With International Help	&  Finland votes: Stubb wins presidency & 	Already two ufo references in Super Bowl commercials lol \\
socialism & Stopping the Cop Cities Countrywide &  Historic Newton Teachers Strike Highlights Divided MA Democratic Party & Biden Docs Confirm Hunter's Pay-To-Play Was A Family Affair\\
politics & Trump Asks Supreme Court to Pause Ruling Denying Him Absolute Immunity& Right-wing judges flaunting their bias and conflicts threaten democracy& Finland votes: Stubb wins presidency\\
unitedkingdom & Sinn Féin politician secretly attended son's PSNI graduation & HMS Prince of Wales sailed today to participate in NATO's Exercise Steadfast Defender & Mongolia Commits to Fighting Corruption With International Help\\
\bottomrule
    \end{tabular}
    \caption{Examples of post titles used in the knowledge quizzes. Participants must correctly select the post titles that belong to the subreddit (i.e., those in columns ``Title 1'' and ``Title 2'').}
    \label{tab:kq}
    \Description[Post titles used in the knowledge quizzes]{
    The table shows post titles used in the knowledge quizzes for four subreddits --- r/worldnews, r/socialism, r/politics, and r/unitedkingdom. There are three post titles for each subreddit.
    }
\end{table*}

\subsection{Survey Details}
Participants were compensated a prorated \$15/hr for completing the survey with a total of \$1,070 spent (including pilot studies, screenings, etc.). 

We provide the survey instrument used in the study as follows:
\begin{enumerate}
    \item In what year did you create your Reddit account?
\item Typically, how often do you post or comment on Reddit versus browsing what others have submitted (lurking)? 
\item Select up to 3 subreddits that you are an active member of. 
\item Are you an active member of any other subreddits that focus on political or social issues outside of those you have already selected? 
\item Provide the name of a subreddit that you are an active member of which focuses on political or social issues that is not one of the subreddits you have already selected. 
\item How diverse are the people for each subreddit? 
\item How included and able to contribute do new and existing members feel for each subreddit? 
\item Select up to 5 topics that are relevant to the subreddit. 
\item Which viewpoint on the topic do you agree with more? 
\item Indicate your level of agreement with the viewpoint. 
\item Indicate the level of agreement you think the majority of subreddit members have for the viewpoint.
\item Disregarding your own stance on the topic, should members of the subreddit be able to share this viewpoint? 
\item Imagine that you are browsing posts, and you see a post related to the selected topic. You notice that in the comments the following viewpoint has not been raised yet. Rate the likelihood you would share this viewpoint on the subreddit under your main account. 
\item Rate the likelihood you would upvote someone else’s comment expressing this viewpoint, if the comment was present.
\end{enumerate}

\subsection{Descriptive Statistics}
\label{subsec:app_descriptive}
\begin{figure}
    \centering
    \includegraphics[width=0.9\linewidth]{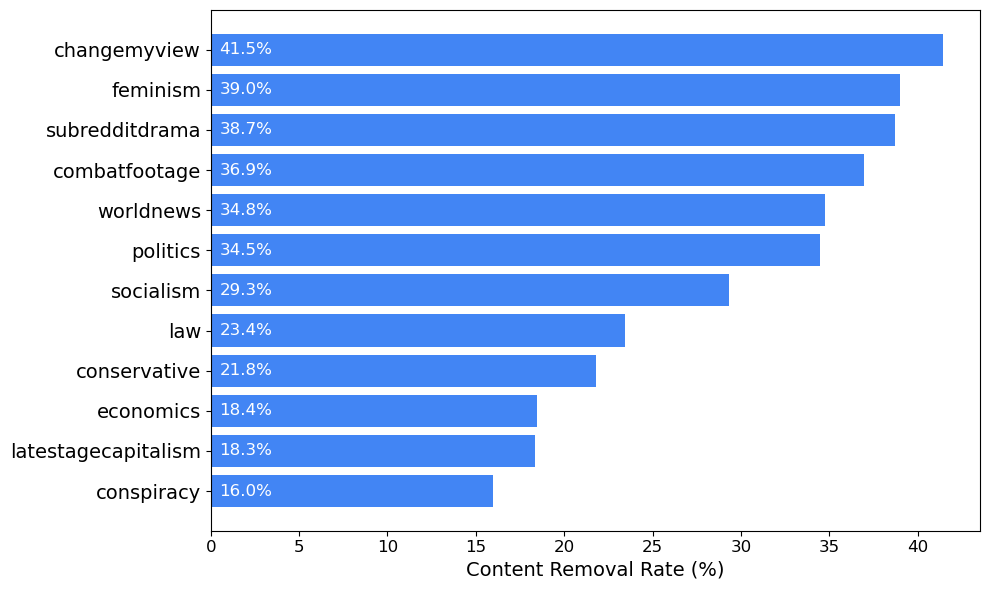}
    \caption{Disaggregated content removal rates for the twelve subreddits included in our analyses. Content removal rates range from $16.0\%$ for r/conspiracy to $41.5\%$ for r/changemyview. }
    \label{fig:disagg}
    \Description[Bar chart of content removal rates across subreddits.]{
    Bar chart of content removal rates across subreddits. The figure ranks subreddits by the percentage of content removed by moderators. changemyview has the highest removal rate at 41.5\%, followed by feminism (39.0\%) and subredditdrama (38.7\%). Lower rates are observed in economics (18.4\%), latestagecapitalism (18.3\%), and conspiracy (16.0\%). Overall, the figure highlights substantial variation in moderation intensity across communities, with some subreddits removing over twice as much content as others.
    }
\end{figure}

Finally, we provide the descriptive statistics of the variables we use in our linear mixed-effect models in Table~\ref{tab:variables} and the disaggregated content removal rates by subreddit (Fig.~\ref{fig:disagg}). 

\begin{table*}[]
    \small
    \centering
    \caption{An overview of descriptive statistics for all introduced variables (min, max, mean, median).}
    \begin{tabular}{p{1.75in}p{1.3in}rrrr}
    \toprule
     &  Variable  & Min & Max & Mean & Median\\
    \midrule \\  [-1.2ex]
    \multirow{2}{1.5in}{Dependent Variables} & Share Likelihood & 1 & 7 & 3.87 & 4\\
    & Upvote Likelihood & 1 & 7 & 5.47 & 6\\[1.2ex]
    \hline \\ [-1.2ex]
    \multirow{7}{0.75in}{Controls} 
    & WTSC & 1.5 & 5.5 & 3.68 & 3.63\\
    & Agreement Intensity & 1 & 3 & 2.17 & 2\\
    & Male (0/1) & 0 & 1 & 0.73 & 1 \\
    & White (0/1) & 0 & 1 & 0.67 & 1\\ 
    & Democrat (0/1) & 0 & 1 & 0.37 & 0\\
    & Account Tenure & 0 & 16 & 6.16 & 6\\ 
    & Posting & 1 & 4 & 2.53 & 2\\[1.2ex]
    \hline 
    \\ [-1.2ex]
    \multirow{5}{1.5in}{Independent Variables} & Incongruency (0/1) & 0 & 1 & 0.25 & 0 \\
    & Inclusion & 1 & 11 & 6.39 & 7\\
    & Diversity & 1 & 11 & 6.33 & 6\\
    & Content Removal Rate (\%) & 15.99 & 41.46 & 30.63 & 34.47\\ 
    \bottomrule
    \end{tabular}
    \label{tab:variables}
    \Description[Descriptive statistics for all study variables. ]{
   The table reports minimum, maximum, mean, and median values for dependent, control, and independent variables.
    }
\end{table*}

\section{Robustness Checks}
\label{sec:robustness}
In this section, we conduct robustness checks on our analyses. First, we present comparisons of our dataset post-filtering with the unfiltered dataset. Second, we show results using an alternative dichotomization of \textit{Incongruency}. Then, we report using results with \textit{ModRatio} for moderation. In our pre-registration, we planned on using both \textit{ModRatio} and \textit{Content Removal Rate} as measures of community moderation, but ultimately removed \textit{ModRatio} as the two were correlated. Then, we provide results for our analysis, including a quadratic term for \textit{Content Removal Rate}. Finally, we report results using an ordinal mixed-effects model. 

\subsection{Comparing filtered dataset }
\label{sec:app_comparing}
\begin{table*}[!htbp]
\centering
\begin{tabular}{llrrlr}
\toprule
\textbf{Variable} & \textbf{Test} & \textbf{Filtered Mean} & \textbf{Unfiltered Mean} & \textbf{Test Statistic} & \textbf{p-value} \\
\midrule
Comment Likelihood & Welch's t-test & 3.88 & 3.79 & t(926.4)=-0.67 & 0.51 \\
Upvote Likelihood & Welch's t-test & 5.47 & 5.40 & t(936.0)=-0.65 & 0.52 \\
\hline 
WTSC & Welch's t-test & 3.68 & 3.68 & t(934.7)=-0.01 & 0.99 \\
Agreement Intensity & Welch's t-test & 6.17 & 6.14 & t(950.9)=-0.62 & 0.54 \\
Gender & Chi-squared    & ---    & ---    & $\chi^2(2) = 0.01$ & 0.99\\
White (0/1) & Chi-squared    & ---    & ---    & $\chi^2(1) = 0.00$ & 0.98 \\ 
Political Affiliation & Chi-squared    & ---    & ---    & $\chi^2(4) = 0.01$ & 0.99 \\
Account Tenure & Welch's t-test & 6.16 & 5.78 & t(928.2)=-1.63 & 0.10 \\
Posting & Welch's t-test & 2.53 & 2.51 & t(948.5)=-0.40 & 0.69 \\
\hline
Incongruency & Chi-squared    & ---    & ---    & $\chi^2(1) = 0.00$ & 0.98 \\
Diversity & Welch's t-test & 5.33 & 5.31 & t(957.6)=-0.19 & 0.85 \\
Inclusion & Welch's t-test & 5.39 & 5.37 & t(948.6)=-0.14 & 0.89 \\
Content Removal Rate & Welch's t-test & 0.31 & 0.30 & t(993.9)=-1.65 & 0.10 \\
\bottomrule
\end{tabular}
\\[1em]
\caption{We compare the filtered and unfiltered data, conducting t-test and $\chi^2$ differences for our independent and dependent variables. There are no statistically significant differences between our filtered and unfiltered data.}
\label{table: comparison}
\end{table*}

We compare the filtered dataset that we use in our analyses with the full, unfiltered dataset to examine if there are any participants or subreddits that are systematically excluded. As shown in Table~\ref{table: comparison}, we conducted statistical analyses comparing the distributions of key variables between the full dataset and the filtered sample. Across all tests, we fail to reject the null hypothesis that filtered responses come from the same distribution as unfiltered responses. While there are no statistically significant differences between the filtered and unfiltered dataset, we do note the two following differences. First, participants in the filtered dataset have longer account tenures ($6.16$ years in the filtered dataset compared to $5.78$ for the unfiltered case). Second, subreddits in the filtered dataset have slightly higher content remove rates with a mean of $30.6\%$ compared to $29.8\%$, although this difference is not statistically significant ($t(993.9) = -1.65, p=0.10$). 

\subsection{Using an alternative operationalization of incongruency}
\label{sec:app_altinc}
\begin{table*}[]
    \small
    \centering
    \caption{Our results are robust to alternative dichotomizations of opinion incongruency. We present the results of a linear mixed-effect model that predicts a participant's likelihood of sharing their opinion to a subreddit using $\IncongruencyAlt$, which binarizes opinion incongruency based off a continuous score, as an independent variable and results after adding interaction effects.}
    \begin{tabular}{lrrlrrl}
    \toprule
         & \multicolumn{3}{c}{$\IncongruencyAlt$} & \multicolumn{3}{c}{$\IncongruencyAlt$ \text{w/ Interactions}}\\
    \cmidrule(lr){2-4} \cmidrule(lr){5-7}
     & Coef. & SE & $p$ & Coef. & SE & $p$ \\ \midrule
    \textit{Fixed Effects} & & \\
    (Intercept) & $4.63$ & $0.59$ & $0.00^{***}$ & $4.64$ & $0.59$ & $0.00^{***}$ \\ \addlinespace
    
    \textbf{Controls} \\ 
    WTSC & $-0.28$ & $0.25$ & $0.27$ & $-0.29$ & $0.25$ & $0.25$ \\ 
    Agreement Intensity & $0.50$ & $0.06$ & $0.00^{***}$ & $0.50$ & $0.06$ & $0.00^{***}$ \\
    Male (0/1) & $-0.72$ & $0.55$ & $0.19$ & $-0.72$ & $0.55$ & $0.20$ \\
    White (0/1) & $0.07$ & $0.53$ & $0.89$ & $0.07$ & $0.53$ & $0.90$ \\ 
    Democrat (0/1) & $-0.71$ & $0.49$ & $0.16$ & $-0.71$ & $0.50$ & $0.16$ \\
    Account Tenure & $0.08$ & $0.24$ & $0.74$ & $0.09$ & $0.24$ & $0.71$ \\
    Posting & $0.68$ & $0.24$ & $0.01^{**}$ & $0.68$ & $0.24$ & $0.01^{**}$ \\ \addlinespace
    
    \textbf{Independent Variables} \\ 
    $\IncongruencyAltNorm$ (0/1) & $-1.21$ & $0.40$ & $0.00^{**}$ & $-1.45$ & $0.70$ & $0.04^{*}$ \\ 
    Inclusion & $0.09$ & $0.10$ & $0.37$ & $0.10$ & $0.10$ & $0.32$ \\
    Diversity & $-0.01$ & $0.10$ & $0.93$ & $-0.04$ & $0.09$ & $0.64$ \\
    Content Removal Rate & $-0.08$ & $0.08$ & $0.35$ & $-0.06$ & $0.08$ & $0.48$ \\ \addlinespace
    
    $\IncongruencyAltNorm$ $\times$ Inclusion &  &  &  & $-0.53$ & $0.56$ & $0.35$ \\
    $\IncongruencyAltNorm$ $\times$ Diversity &  &  &  & $0.78$ & $0.39$ & $0.04^{*}$ \\
    $\IncongruencyAltNorm$ $\times$ Content Removal Rate &  &  &  & $-0.97$ & $0.51$ & $0.06$ \\
    [1.2ex]
    \hline
    \textit{Model Fit} & & \\ 
    Marginal $R^2$ & $0.232$ & & & $0.236$ & \\
    Conditional $R^2$ & $0.798$ & & & $0.797$ & \\
    \hline \\[-1.8ex] 
    \textit{Note: $^{*}$p$<$0.05; $^{**}$p$<$0.01; $^{***}$p$<$0.001} \\ 
    \bottomrule
    \end{tabular}
    \label{tab:alt_inc}
\end{table*}
We demonstrate that our findings are robust to different definitions of our measurement for opinion incongruency. In Sec.~\ref{sec:measures}, we define \textit{Incongruency} as 1 if participants agree with the viewpoint and the majority of subreddit members are neutral or disagree; \textit{Incongruency} is 0 if the participant agrees and the majority of the subreddit also agrees. In, our alternative dichotomization, $\mathit{Incongruency}_{\mathrm{Alt}}$, we compute the absolute difference between how much the participant agrees with a viewpoint, \textit{Agreement}, and how much they believe the majority of subreddit members agree with the viewpoint. The continuous score ranges from $0$ to $7$. Absolute differences greater than $4$ are coded as 1 (i.e., incongruent) and less than or equal to $4$ are coded as 0. The correlation (Pearson's $r$) between \textit{Incongruency} and $\IncongruencyAlt$ is $0.288$. Using \textit{Incongruency} results in a slightly better model fit than $\IncongruencyAlt$, as indicated by its lower AIC score ($1462.7$ vs. $1489.2$).

As shown in Table~\ref{tab:alt_inc}, $\IncongruencyAlt$ has a negative relationship with the likelihood of sharing ($\beta=-1.21$, $p=0.002$). Furthermore, we observe qualitatively similar results with the interaction effects between $\IncongruencyAlt$ and \textit{Diversity} ($\beta=0.78$, $p=0.044$) as well as $\IncongruencyAlt$ and \textit{Content Removal Rate} ($\beta=-0.97$, $p=0.059$).

\subsection{Replacing measures for moderation} 
\label{sec:app_mod}
\begin{table*}[]
    \small
    \centering
    \caption{Including \emph{ModRatio} does not substantially alter relationships between our independent variables and the likelihood of sharing a viewpoint. The results of a linear mixed-effect model that predicts a participant's likelihood of sharing their opinion to a subreddit. We present results including \emph{ModRatio} as an independent variable and results after adding interaction effects.}
    \begin{tabular}{lrrlrrl}
    \toprule
         & \multicolumn{3}{c}{ModRatio} & \multicolumn{3}{c}{ModRatio w/ Interactions}\\
    \cmidrule(lr){2-4} \cmidrule(lr){5-7}
     & Coef. & SE & $p$ & Coef. & SE & $p$ \\ \midrule
    \textit{Fixed Effects} & & \\
    (Intercept) & $4.75$ & $0.59$ & $0.00^{***}$ & $4.81$ & $0.58$ & $0.00^{***}$\\  \addlinespace
    \textbf{Controls} \\
    WTSC & $-0.30$ & $0.25$ & $0.22$ & $-0.30$ & $0.25$ & $0.22$ \\ 
    Agreement Intensity & $0.40$ & $0.06$ & $0.00^{***}$ & $0.41$ & $0.06$ & $0.00^{***}$\\
    Male (0/1) & $-0.71$ & $0.55$ & $0.20$ & $-0.83$ & $0.54$ & $0.13$ \\
    White (0/1) & $0.11$ & $0.53$ & $0.83$ & $0.19$ & $0.53$ & $0.72$ \\ 
    Democrat (0/1) & $-0.75$ & $0.49$ &$0.13$ & $-0.75$ & $0.49$ &$0.13$ \\
    Account Tenure &  $0.10$ & $0.24$ & $0.67$ & $0.12$ & $0.24$ & $0.63$ \\
    Posting & $0.64$ & $0.24$ & $0.01^{**}$ & $0.65$ & $0.24$ & $0.01^{**}$\\ \addlinespace
    \textbf{Independent Variables} \\
    Incongruency (0/1) & $-0.74$ & $0.14$ & $0.00^{***}$ & $-0.57$ & $0.15$ & $0.00^{***}$ \\
    Inclusion & $0.07$ & $0.10$ & $0.49$ & $0.03$ & $0.11$ & $0.81$\\
    Diversity & $0.02$ & $0.10$ & $0.84$ & $-0.06$ & $0.10$ & $0.56$\\
    ModRatio & $-0.15$ & $0.10$ & $0.14$ & $-0.07$ & $0.10$ & $0.53$ \\ \addlinespace
    Incongruency $\times$ Inclusion & & & & $0.21$ & $0.14$&$0.14$ \\
    Incongruency $\times$ Diversity & & & & $0.38$ & $0.15$ & $0.01^{*}$\\
    Incongruency $\times$ ModRatio & & & & $-0.35$ & $0.16$ & $0.02^{*}$\\
    [1.2ex]
    \midrule
    \textit{Model Fit} & & \\ 
    Marginal $R^2$ & $0.25$ & & & $0.26$ & \\
    Conditional $R^2$ & $0.81$ & & & $0.81$ & \\
    \hline \\[-1.8ex] 
    \textit{Note: $^{*}$p$<$0.05; $^{**}$p$<$0.01; $^{***}$p$<$0.001} \\ 
    \bottomrule
    \end{tabular}
    \label{tab:modratio}
\end{table*}
We examine changes to the model when replacing \textit{Content Removal Rate} with \textit{ModRatio} as our measure of moderation. \textit{ModRatio} is the ratio of the number of subscribers to number of moderators in a subreddit. In this case, a higher \textit{ModRatio} would mean less active moderation as there are fewer moderators per subscribers. We apply a logarithmic transformation with base 2 after adding a start-value of 1 to \textit{ModRatio}. As shown in Table~\ref{tab:modratio}, there is a significant negative interaction between \textit{Incongruency} and \textit{ModRatio} ($\beta=-0.35$, $p = 0.02$), mirroring our results when using \textit{Content Removal Rates}. The trends with our independent variables are similar to when we used \textit{Content Removal Rate}.

\subsection{Examining quadratic terms for moderation}
\label{sec:app_quad}
\begin{table*}[]
    \small
    \centering
    \caption{The results of a linear mixed-effect model predicting a participant's likelihood of sharing their opinion in a subreddit. We present results including a quadratic term for \emph{Content Removal Rate} as an independent variable.}
    \begin{tabular}{lrrl}
    \toprule
     & Coef. & SE & $p$ \\ \midrule
        \textit{Fixed Effects} & & \\
    (Intercept) & $4.71$ & $0.60$ & $0.00^{***}$\\  \addlinespace
    \textbf{Controls} \\
    WTSC & $-0.30$ & $0.25$ & $0.23$\\ 
    Agreement Intensity & $0.40$ & $0.06$ & $0.00^{***}$\\
    Male (0/1) & $-0.69$ & $0.55$ & $0.22$ \\
    White (0/1) & $0.10$ & $0.53$ & $0.86$ \\ 
    Democrat (0/1) & $-0.74$ & $0.49$ &$0.14$ \\
    Account Tenure &  $0.08$ & $0.24$ & $0.75$ \\
    Posting & $0.71$ & $0.24$ & $0.01^{**}$ \\ \addlinespace
    \textbf{Independent Variables} \\
    Incongruency (0/1) & $-1.09$ & $0.22$ & $0.00^{***}$ \\
    Content Removal Rate & $0.06$ & $0.12$ & $0.62$\\ 
    $\text{Content Removal Rate}^2$ & $0.04$ & $0.11$ & $0.70$\\
    Incongruency $\times \text{Content Removal Rate}$ & $-0.02$ & $0.19$ & $0.90$\\ 
    Incongruency $\times \text{Content Removal Rate}^2$ & $0.32$ & $0.18$ & $0.07$\\
    \addlinespace
    [1.2ex]
    \midrule
    \textit{Model Fit} & & \\ 
    Marginal $R^2$ & $0.25$  \\
    Conditional $R^2$ & $0.81$ \\
    \hline \\[-1.8ex] 
    \textit{Note: $^{*}$p$<$0.05; $^{**}$p$<$0.01; $^{***}$p$<$0.001} \\ 
    \bottomrule
    \end{tabular}
    \label{tab:quadratic}
\end{table*}
We also report results of our linear mixed-effects model after including a quadratic term for \textit{Content Removal Rate} in Table~\ref{tab:quadratic}. Again, we see similar trends between \textit{Incongruency} and \textit{Share Likelihood}. There is no significant relationship between the quadratic term and our dependent variable. 

\subsection{Treating sharing likelihood as ordinal}
In Table~\ref{tab:model1_ordinal}, we use an ordinal mixed-effect regression to predict \textit{Share Likelihood}, which is a Likert-scale with values ranging from $1$ to $7$. In the main body of the paper, we report results treating \textit{Share Likelihood} as continuous, as is consistent with other works measuring the spiral of silence~\cite{gearhart2014gay,gearhart2015something,willnat2002individual,ho2013spiral}. This decision is also consistent with prior work in statistics, demonstrating the appropriateness of such an approach~\cite{sullivan2013analyzing,norman2010likert}. 

Our conclusions about direction are robust across continuous and ordinal specifications; however, statistical significance varies with modeling assumptions. First, we note there is a statistically significant negative relationship between \textit{Incongruency} ($\beta=-0.87$, $p<0.001$ in Model 3b) and the likelihood of sharing a viewpoint. Furthermore, we note a positive interaction effect between \textit{Diversity} and opinion sharing ($\beta=0.26$) as well as a negative interaction effect between \textit{Content Removal Rate} and opinion sharing ($\beta=-0.28$) although the interaction effects are not significant at our level of statistical power ($p=0.11$ for both).
\begin{table*}[]
    \small
    \centering
    \caption{We present results using an ordinal mixed-effect model. The direction of the relationships are robust to the model reported in Sec.~\ref{sec:results}.}
    \begin{tabular}{lrrlrrlrrl}
    \toprule
         & \multicolumn{3}{c}{Model 3a} &  \multicolumn{3}{c}{Model 3b} &  \multicolumn{3}{c}{Model 3c}\\
    \cmidrule(lr){2-4}\cmidrule(lr){5-7}\cmidrule(lr){8-10}
     & Coef. & SE & $p$ & Coef. & SE & $p$ & Coef. & SE & $p$\\ \midrule
    \textbf{Dependent Variable} \\ 
    Share Likelihood $\leq$ 1 & $-2.45$ & $0.68$ & $0.00^{***}$& $-2.84$ & $0.71$ & $0.00^{***}$&$-2.96$ & $0.71$ & $0.00^{***}$ \\
    Share Likelihood $\leq$ 2 & $-1.20$ & $0.68$ & $0.08$ & $-1.53$ & $0.70$ & $0.03^{*}$ & $-1.61$ & $0.70$ & $0.02^{*}$\\
    Share Likelihood $\leq$ 3 & $-0.72$ & $0.68$ &$0.29$& $-1.02$ & $0.70$ & $0.14$ & $-1.10$ & $0.70$ & $0.00^{***}$\\
    Share Likelihood $\leq$ 4 & $-0.17$ & $0.68$ &$0.81$& $-0.46$ & $0.69$ & $0.51$ & $-0.47$ & $-0.52$ & $0.46$\\
    Share Likelihood $\leq$ 5 & $0.83$ & $0.68$ &$0.22$& $0.62$ & $0.69$ & $0.37$ & $0.59$ & $0.70$ & $0.40$\\
    Share Likelihood $\leq$ 6 & $1.97$ & $0.68$ &$0.00^{**}$& $1.83$ & $0.70$ & $0.00^{***}$ & $1.80$ & $0.70$ & $0.01^{**}$\\
    \textbf{Controls} \\
    WTSC & $-0.42$ & $0.29$ & $0.15$ &$-0.46$ &$0.29$ &$0.12$ &$-0.47$ & $0.29$ & $0.11$\\ 
    Agreement Intensity & $0.60$ & $0.08$ & $0.00^{***}$ & $0.53$ & $0.08$ & $0.00^{***}$ & $0.57$ & $0.08$ & $0.00^{***}$\\
    Male (0/1) & $-0.58$ & $0.63$ & $0.35$ & $-0.66$ & $0.64$ & $0.30$ & $-0.70$ & $0.64$ & $0.27$\\
    White (0/1) & $0.10$ & $0.61$ & $0.87$ & $0.11$ & $0.62$ & $0.86$ & $0.11$ & $0.62$ & $0.86$\\ 
    Democrat (0/1) & $-0.77$ & $0.57$ & $0.18$ & $-0.87$ & $0.58$ & $0.13$ & $-0.89$ & $0.58$ & $0.13$\\
    Account Tenure & $0.06$ & $0.27$ & $0.81$ & $0.05$ & $0.28$ & $0.84$ & $0.07$ & $0.28$ & $0.81$\\
    Posting & $0.80$ & $0.28$ & $0.00^{**}$& $0.77$ & $0.29$ & $0.01^{**}$ & $0.80$ & $0.29$ & $0.01^{**}$\\ \addlinespace
    \textbf{Independent Variables} \\
    Incongruency (0/1) & & & & $-0.87$ & $0.17$ & $0.00^{***}$ & $-0.80$ & $0.17$& $0.00^{***}$\\
    Inclusion & & & & $0.10$ & $0.12$ & $0.41$ & $-0.02$ & $0.13$ & $0.88$\\
    Diversity & & & & $0.02$ & $0.11$ & $0.88$ & $-0.04$ & $0.12$ & $0.76$\\
    Content Removal Rate & & & & $-0.07$ & $0.10$ & $0.50$ & $0.00$ & $0.10$ &$0.99$\\ \addlinespace
    Incongruency $\times$ Inclusion  & & & & & & & $0.49$ & $0.19$ & $0.01^{*}$\\
    Incongruency $\times$ Diversity & & & & & & & $0.26$ & $0.17$ & $0.11$\\
    Incongruency $\times$ Content Removal Rate & & & & & & & $-0.28$ & $0.17$ & $0.11$\\
    [1.2ex]
    \hline \\[-1.8ex] 
    \textit{Note: $^{*}$p$<$0.05; $^{**}$p$<$0.01; $^{***}$p$<$0.001} \\ 
    \bottomrule
    \end{tabular}
    \label{tab:model1_ordinal}
\end{table*}

\section{Deviations from Pre-registration}
We pre-register our analyses on OSF. We also report the following deviations from our pre-registration. First, we change our statistical model to include a crossed random effect between subreddit and participant, as we are capturing multiple observations per subreddit per participant. We updated our power analysis with this updated model to reach our target sample size of 270 responses. Second, we split community values into two separate hypotheses, as we found perceived inclusion and diversity were measuring distinct constructs. We also updated the wording on our hypotheses to mention incongruent viewpoints. Third, we changed how we measured \textit{Content Removal Rate}. In the pre-registration, we had planned to use 500 posts sampled from each subreddit; however, during analysis, we found this method underreported the amount of content removal compared to results from prior work~\cite{jhaver2019does}. Thus, we decided to rely on Pushshift data~\cite{pushshift2023dumps}, which gave us access to a larger corpus of posts. Fourth, we remove the variable \textit{Mod Ratio} from our models since it was correlated with \textit{Content Removal Rate}. For robustness, we replicate our results using \textit{Mod Ratio} as our measure for community moderation, finding similar results. Finally, we conducted additional analyses using \textit{Upvote Likelihood} as our dependent variable, which are presented as post-hoc analyses.

\section{Other Techniques for Avoiding Self-Silencing}
In our survey, we also ask participants what other actions they are likely to take for viewpoints they would not share under their main account. Participants (N=68) are most likely to ``lurk,'' or read discussion on a topic but not share their viewpoint. After lurking, the second most common option is to discuss the viewpoint offline with friends and family (N=34). Prior work~\cite{pewresearchcenter_2014_sos} found that people are more likely to engage in conversation on controversial political topics in face-to-face settings rather than on social media platforms. 
\end{document}